\documentclass[aps,preprint,tightenlines,floatfix,groupedaddress,nofootinbib]{revtex4}

\usepackage{graphicx}
\usepackage{dcolumn}

\begin{document}
\bibliographystyle{apsrev}

\preprint{\vbox {\hbox{BNL-HET-05/25} \hbox{WSU--HEP--0506} \hbox{hep-ph/0510274}}}

\vspace*{2cm}

\title{\boldmath Neutrinos in a left-right model with a horizontal symmetry}

\author{Ken Kiers}
\email{knkiers@taylor.edu}
\author{Michael Assis}{\bf }
\email{michael_assis@taylor.edu}
\author{David Simons}
\email{dlsimons@bcm.edu}
\altaffiliation{Current Address: Baylor College of Medicine, One Baylor Plaza, Houston, TX 77030}
\affiliation{Physics Department, Taylor University, 
236 West Reade Ave., Upland, Indiana 46989}

\author{Alexey A. Petrov}
\email{apetrov@physics.wayne.edu}
\affiliation{Department of Physics and Astronomy, Wayne State University,
Detroit, Michigan 48201}

\author{Amarjit Soni}
\email{soni@bnl.gov}
\affiliation{High Energy Theory, Department of Physics,
Brookhaven National Laboratory, Upton, New York 11973-5203}

\date{\today}

\begin{abstract}
We analyze the lepton sector of a Left-Right Model based
on the gauge group $SU(2)_L\times SU(2)_R\times U(1)$, concentrating mainly on 
neutrino properties. 
Using the seesaw mechanism and a horizontal symmetry, we keep the 
right-handed symmetry breaking scale relatively low,
while simultaneously satisfying phenomenological constraints
on the light neutrino masses.
We take the right-handed scale to be of order 10's of TeV and
perform a full numerical analysis of the model's parameter space, subject
to experimental constraints on neutrino masses and mixings.
The numerical procedure yields results for the right-handed neutrino masses and mixings
and the various CP-violating phases.
We also discuss phenomenological applications of the model to
neutrinoless double beta decay,
lepton-flavor-violating decays (including decays such as $\tau \to 3\mu$) and leptogenesis.
\end{abstract}


\maketitle

\section{Introduction} \label{intro}

The Standard Model (SM) of particle physics has thus far provided an incredibly accurate
description of all observed experimental data. Nevertheless, the SM is widely regarded as 
being a low-energy effective theory, with a limited range of applicability
and predictability. New interactions must arrive at the energy scale of several TeV 
in order to explain such features of the SM as the quark and lepton mass hierarchy. 

An intriguing aspect of the SM is that its weak interaction sector represents the 
only known interaction that distinguishes between the right- and left-handed 
fermions. This aspect is addressed in a set of aesthetically-pleasing ``left-right models'' 
(LRMs) based on the gauge group $SU(2)_L\times SU(2)_R\times U(1)$, where the left- and 
right-handed fermion fields are treated symmetrically~\cite{patisalam1974,mohapatrapati,
mohapatrapati1975,senjanovic,mohapatrapaige,senjanovic1979,duka}. In such models,
left-right symmetry is broken at some high scale, yielding a parity-violating Standard 
Model-like theory at low energies. 
In typical phenomenological studies of the LRM, the right-handed scale (i.e., the scale at which
$SU(2)_R$ is broken) is assumed to be very high, of the 
order of $10^{10}$~GeV. Such a high energy scale would render direct experimental verification 
of a LR-motivated scenario impossible in the near future.  By way of contrast, more 
moderate values for the right-handed scale -- in the range 20-50 TeV, say -- could 
have observable consequences at experiments in the near future~\cite{ball1,kiersLR}.  
This is the energy range that we shall consider in this work.

Recently, there has been a renewed interest in the LRM, in particular, due to 
the discovery of neutrino oscillations and to
major advances in 
experimental studies of CP-violation in the quark sector.  Of interest to us in this work is that
the LRM provides a natural process for the suppression of neutrino
masses through the seesaw mechanism.  LRMs also offer additional sources of CP violation,
coming both from the
right-handed Cabibbo-Kobayashi-Maskawa (CKM) and Maki-Nakagawa-Sakata (MNS) 
matrices as well as from the Higgs sector of the theory~\cite{kiers_HiggsLR}.
The CP violation occuring in the leptonic sector of the model could in principle
be of interest within the context of leptogenesis~\cite{Fukugita:1986hr,Luty:1992un,
Buchmuller:1996pa,Hambye:2003ka,Antusch:2004xy,Chen:2004ww,Atwood:2005bf}.
It is important to emphasize that the phases in the left-handed MNS matrix are 
currently unconstrained.

As noted above, the LRM contains the natural possibility of implementing
the seesaw mechanism for the generation of small neutrino masses.
In the simplest version of the seesaw mechanism, only the SM-singlet right-handed neutrinos
are initially allowed to obtain Majorana mass terms, resulting in the neutrino mass 
matrix
\begin{eqnarray} \label{eq:StdCSaw}
        {\cal M} = \left(\begin{array}{cc}
                        0 & M_{LR}\\
                        M_{LR}^T & M_{RR}\\
                        \end{array}\right) \; ,
		\label{eq:noMLL}
\end{eqnarray}
where $M_{RR}$ and $M_{LR}$ are Majorana and Dirac mass matrices, respectively,
in flavour-space.
This construction, with a block of ``zeros'' where the left-handed Majorana mass terms would go,
leads to what is sometimes called the ``Type I'' seesaw mechanism.  It can be generated 
in the LRM.
An approximate block-diagonalization of Eq.~(\ref{eq:noMLL}), assuming that 
the elements in $M_{RR}$ are much larger than those in $M_{LR}$, leads to the standard 
seesaw expression for the light neutrino mass 
matrix, $M_\nu \simeq - M_{LR} M_{RR}^{-1} M_{LR}^T\equiv M_\nu^{I}$. 

To implement the seesaw mechanism in the LRM, one introduces a right-handed Higgs triplet field, 
$\Delta_R$, into the theory.
The field $\Delta_R$ serves a dual purpose -- it breaks the left-right symmetry of the model at a high scale
and it also couples to Majorana neutrino fields, giving rise to the right-handed Majorana mass matrix $M_{RR}$
required for the seesaw mechanism.  $M_{RR}$ is proportional to
the right-handed symmetry breaking scale, making it naturally large.
Left-right symmetry also requires the existence of a left-handed Higgs triplet field, $\Delta_L$.
The neutral components of the left- and right-handed triplet fields both 
generically obtain a vacuum expectation value (VEV) under spontaneous
symmetry breaking, $\langle\delta^0_{L,R}\rangle= v_{L,R}e^{i\theta_{L,R}}/\sqrt{2}$.
LRMs also typically contain a bidoublet field, $\phi$, with VEVs at the weak scale.  The role
played by $\phi$ is similar to that played by the usual Higgs doublet of the SM.

In contrast with the situation in the Type I seesaw mechanism,
LRMs generically also contain a non-zero left-handed Majorana mass matrix for neutrinos, $M_{LL}$.
This mass matrix is proportional to $v_L$ (the VEV associated with the left-handed Higgs triplet)
and $v_L$ need not be zero.  There is, in fact, a seesaw-like relation among the VEVs $v_L$ and $v_R$
in the LRM that arises due to terms in the Higgs potential that couple $\phi$, $\Delta_L$ and $\Delta_R$
($\mbox{Tr}(\phi\Delta_R\phi^\dagger\Delta_L^\dagger)$, for example).
If all dimensionless coefficients in the Higgs potential are of order unity, one finds that
$v_L\sim k^2/v_R$, where $k$ is a dimensionful quantity of order the weak scale~\cite{Mohapatra:1980yp,
deshpande,kiers_HiggsLR}.  The situation $v_L=0$ (and hence $M_{LL}=0$) can be obtained by dropping the offending
terms from the Higgs potential, although attempts to disallow the terms using a symmetry
meet with difficulties~\cite{deshpande}.
Inclusion of a non-zero matrix $M_{LL}^\dagger$ in the ``zero'' block of Eq.~(\ref{eq:noMLL})
(see Eq.~(\ref{eq:6by6}) below)
leads to the ``Type II'' seesaw mechanism~\cite{Mohapatra:1980yp,
Mohapatra:2000qe}, 
yielding the following approximate mass matrix
for light neutrinos,
\begin{equation}
	M_\nu \simeq M_{LL}^\dagger - M_{LR} M_{RR}^{-1} M_{LR}^T \equiv M_\nu^{II}+M_\nu^{I},
		\label{eq:mnu1}
\end{equation}
with $M_{LL}$ generically of order $v_L \sim k^2/v_R$ and $M_{RR}\sim v_R$.  
Experimentally, the terms in $M_\nu$ must be at most of order about 0.1~eV.
If the largest terms in $M_{LR}$ are taken to be of order $m_\tau$,
and if $M_{LL}$ does not undergo any further suppression,
then the first term in Eq.~(\ref{eq:mnu1}) 
dominates and sets the minimum scale $v_R$ that is phenomenologically viable.
Taking $k$ to be of order the weak scale we find that $v_R$ would need to be at least of order $10^{14}$~GeV
in this case, ruling out any possibility of observing LR-induced effects at collider experiments.
The analogous lower bound coming from the second term alone (Type I seesaw) is of order $10^{10}$~GeV
if one assumes $M_{LR}\sim m_\tau$.

A few approaches have been suggested for reducing the right-handed scale $v_R$ while
simultaneously satisfying phenomenological constraints within the context of Type II models.
One approach is to suppress $M_{LL}$ by separating the 
parity and gauge symmetry breaking scales. For instance, one can introduce a 
pseudoscalar Higgs field $\eta$ that acquires a vacuum expectation value (thereby breaking parity) 
at a very high scale, while allowing the right-handed gauge symmetry to be broken at a much lower 
scale~\cite{Chang:fu,Chang:1985en}
This approach could lead to interesting phenomenology. In such a model,
$M_{LL}\sim k^2v_R/\langle\eta\rangle^2$, so that the left-handed Majorana mass terms
could be of an acceptable size provided that 
$\langle\eta\rangle$ were sufficiently high.  

Another promising approach for bringing $v_R$ down to a potentially observable scale
is to introduce an extra $U(1)$
horizontal symmetry that is broken by a small parameter $\epsilon$~\cite{froggatt, nir1993,khasanov,
kiers_HiggsLR}.  In this approach, each field in the model is assigned a charge under the horizontal symmetry.
Yukawa couplings and the dimensionless coefficients in the Higgs potential are then suppressed by various powers
of $\epsilon$, with the powers depending on combinations of charge assignments.
In addition to providing a nice dynamical mechanism for producing the observed hierarchies in the charged lepton
masses, this approach also allows for a significant reduction in the right-handed scale $v_R$.
As shown in Refs.~\cite{khasanov,kiers_HiggsLR}, an appropriate choice of charge assignments 
leads to a suppression of $v_L$, thereby suppressing $M_{LL}$.  There is a similar suppression of 
the Yukawa couplings involved in $M_{LR}$, the net effect of which is to loosen the stringent lower bounds on $v_R$.

In this work we perform an indepth numerical study of neutrinos 
in a left-right model that is supplemented by such a broken $U(1)$ horizontal
symmetry.
The goal of the paper is to determine the parameter
space of the model already constrained by the neutrino mass and mixing measurements. 
We focus on the case in which the right-handed symmetry breaking scale is only ``moderately'' 
large (20-50~TeV).  In the horizontal symmetry scheme,
right-handed scales as low as 20~TeV can in fact
lead to results consistent with neutrino phenomenology.
The numerical procedure also yields results for the right-handed neutrino masses and mixings
and the various CP-violating phases.
Throughout this work we assume that there are three generations of light neutrinos and
ignore the possibility of light sterile neutrinos.  We also assume the ``normal'' ordering of neutrino masses,
$m_{\nu_1}<m_{\nu_2}<m_{\nu_3}$.

The outline of the remainder of the paper is as follows.  In Sec.~\ref{sec:model} we describe
the LRM and establish our notation.  Section~\ref{sec:numerical} contains our numerical results for the allowed 
parameter space of the model, as well as a discussion of some phenomenological applications such as lepton-flavor 
violating transitions and leptogenesis. We conclude with a brief discussion in Sec.~\ref{sec:conclusions}.
The Appendix contains approximate expressions for the right-handed neutrino
masses and mixings as well as a particular case study.

\section{The Model}
\label{sec:model}

\subsection{Mass and MNS matrices}

The LRM is based on the gauge group $SU(2)_L\times
SU(2)_R\times U(1)_{B-L}$.
We will consider a minimal version of the model, whose
Higgs sector contains a bidoublet Higgs boson field $\phi\sim(2,\overline{2},0)$
and two triplet Higgs boson
fields $\Delta_{L}\sim(3,1,2)$ and $\Delta_{R}\sim(1,3,2)$.
As noted in the Introduction, 
$\Delta_R$ is used to break the gauge symmetry down
to $SU(2)_L\times U(1)_{Y}$ and $\phi$ is used to break it down
to $U(1)_\textrm{\scriptsize{em}}$. 
The various Higgs boson fields may be parametrized as follows, 
\begin{eqnarray}
	\phi = \left(\begin{array}{cc}
		\phi_1^0 & \phi_1^+ \\
		\phi_2^- & \phi_2^0 \\
		\end{array}\right)\; ,~~
	\Delta_{L,R} = \left(\begin{array}{cc}
		\delta^+_{L,R}/\sqrt{2} & \delta^{++}_{L,R} \\
		\delta^0_{L,R} & -\delta^+_{L,R}/\sqrt{2} \\
		\end{array}\right) \; .
\end{eqnarray}
The left- and right-handed lepton fields transform as doublets under 
$SU(2)_L$ and $SU(2)_R$, respectively, and are given by
\begin{eqnarray}
	\psi^\prime_{iL,R} = \left(\begin{array}{c}
		\nu^\prime_{iL,R}\\
		e^\prime_{iL,R}\\
		\end{array}\right) \;,
\end{eqnarray}	
where $i$ is a generation index and where the primes denote
that the fields are gauge eigenstates.  In addition
to the gauge symmetry, it is common to impose an extra
left-right parity symmetry~\cite{deshpande}, demanding invariance under
\begin{eqnarray}
	\psi^\prime_{iL}\leftrightarrow\psi^\prime_{iR},
	~~~\phi\leftrightarrow \phi^\dagger,
	~~~\Delta_L \leftrightarrow \Delta_R.
	\label{eq:parity}
\end{eqnarray}
The lepton Yukawa couplings that
are consistent with the gauge and parity
symmetries discussed above are~\cite{deshpande}
\begin{eqnarray}
	-{\mathcal L}_\textrm{\scriptsize Yukawa} = 
		\overline{\psi}_{iL}^\prime\left(
		G_{ij} \phi + H_{ij} 
		\widetilde{\phi}\right)\psi_{jR}^\prime +
		\frac{i}{2}F_{ij}\left(\psi^{\prime T}_{iL} C \tau_2
		\Delta_L\psi^{\prime}_{jL} +
		\psi^{\prime T}_{iR} C \tau_2
		\Delta_R\psi^{\prime}_{jR}\right)
		+ \textrm{h.c.} \; ,
	\label{eq:yuk}
\end{eqnarray}
where $\widetilde{\phi}= \tau_2 \phi^* \tau_2$ and where
$C=i\gamma^2\gamma^0$ is the charge conjugation matrix.  
In order to respect the parity symmetry in Eq.~(\ref{eq:parity}),
the $3\times 3$ matrices $G$ and $H$ must be Hermitian.
The matrix $F$ is complex, in general, but may be taken to be symmetric
without any loss in generality.\footnote{This follows because one may
write, for example,
$\overline{\nu_{iL}^{\prime c}}\nu_{jL}^{\prime}=\overline{\nu_{jL}^{\prime c}}
\nu_{iL}^{\prime}$,
where $\nu_{iL}^{\prime c} = C\overline{\nu_{iL}^\prime}^T$~\cite{kayser}.}

Upon spontaneous symmetry breaking the Higgs boson fields acquire VEVs, which
may be parametrized as
\begin{eqnarray}
	\langle \phi\rangle =\left(\begin{array}{cc}
		k_1/\sqrt{2} & 0 \\
		0& k_2e^{i\alpha}/\sqrt{2} \\
		\end{array}\right) ,~~~
	\langle \Delta_L \rangle= \left(\begin{array}{cc}
		0 & 0 \\
		v_Le^{i\theta_L}/\sqrt{2} & 0\\
		\end{array}\right),~~~
	\langle \Delta_R \rangle= \left(\begin{array}{cc}
		0 & 0 \\
		v_R/\sqrt{2} & 0\\
		\end{array}\right),
	\label{eq:vevs}
\end{eqnarray}
with $k_{1,2}$ and $v_{L,R}$ real and positive.  Gauge rotations have been used to eliminate possible phases
associated with $k_1$ and $v_R$~\cite{deshpande}.
Phenomenological constraints require that
$v_R\gg k_1, k_2\gg v_L$.  In this case
$k_1$ and $k_2$ satisfy the constraint~\cite{langacker}
\begin{eqnarray}
	k_1^2+k_2^2 \simeq 
		\frac{4m_W^2}{g^2}\simeq(246.2~\textrm{GeV})^2 \; .
		\label{eq:kkpr1}
\end{eqnarray}
Also, it is natural to assume $k_2/k_1\sim m_b/m_t$; in
the numerical work below we shall set $k_2/k_1=3/181$ as in Ref.~\cite{kiersLR}.

The VEVs in Eq.~(\ref{eq:vevs}) lead to Dirac mass terms for 
the neutrinos and charged leptons,
\begin{eqnarray}
	-{\mathcal L}_\textrm{\scriptsize Dirac} = 
		\frac{1}{\sqrt{2}}\;\overline{\nu^\prime_L}\left(Gk_1+H k_2e^{-i\alpha}
		\right)\nu_R^\prime +
		\frac{1}{\sqrt{2}}\;\overline{e^\prime_L}\left(Gk_2e^{i\alpha}+H k_1
		\right)e_R^\prime + \textrm{h.c.} \; ,
	\label{eq:diracmass}
\end{eqnarray}
as well as Majorana mass terms for the neutrinos,
\begin{eqnarray}
	-{\mathcal L}_\textrm{\scriptsize Majorana} = 
		\frac{1}{2\sqrt{2}}\left(\overline{\nu_L^{\prime c}}Fv_Le^{i\theta_L}
			\nu^\prime_L+
			\overline{\nu_R^{\prime c}}Fv_R
			\nu^\prime_R\right)+ \textrm{h.c.},
	\label{eq:majoranamass}
\end{eqnarray}
where $\nu_{L,R}^{\prime c}\equiv C\overline{\nu_{L,R}^\prime}^T$.
The mass matrix for the charged leptons is thus
\begin{eqnarray}
	M_\ell = \frac{1}{\sqrt{2}}\left(Gk_2e^{i\alpha}+H k_1 \right)\; ,
		\label{eq:mlep}
\end{eqnarray}
which may be diagonalized by a biunitary transformation
\begin{eqnarray}
	M_\ell^\textrm{\scriptsize diag} = V_L^{\ell\dagger}
		M_\ell V_R^\ell \; ,
		\label{eq:diagchargedlep}
\end{eqnarray}
where the elements in $M_\ell^\textrm{\scriptsize diag}$ are real
and positive.  

Consideration of the neutrino mass matrix
is slightly complicated by the fact that both Majorana and
Dirac mass terms are present.  Defining
\begin{eqnarray}
	\Psi_\nu^\prime = \left(\begin{array}{c}
			  	\nu_L^\prime \\
				\nu_R^{\prime c} \\
				\end{array}\right) 
\end{eqnarray}
allows the neutrino mass terms in the Lagrangian to be written as
\begin{eqnarray}
	-{\mathcal L}_{m_\nu} = \frac{1}{2}\overline{\Psi_\nu^\prime}
		{\cal M}\Psi_\nu^{\prime c} + \textrm{h.c.},
\end{eqnarray}
where the Majorana and Dirac neutrino mass matrices have been incorporated into
a single $6\times 6$ complex symmetric matrix
\begin{eqnarray}
	{\cal M} = \left(\begin{array}{cc}
			M_{LL}^\dagger & M_{LR}\\
			M_{LR}^T & M_{RR}\\
			\end{array}\right) \; ,
		\label{eq:6by6}
\end{eqnarray}
with
\begin{eqnarray}
	M_{LR}=\frac{1}{\sqrt{2}}\left(Gk_1+H k_2e^{-i\alpha}\right)\; ,~~~~~~
	M_{LL} = \frac{1}{\sqrt{2}}F v_Le^{i\theta_L},~~~~~
	M_{RR} = \frac{1}{\sqrt{2}}F v_R .
		\label{eq:MLRetc}
\end{eqnarray}
The $6\times 6$ neutrino mass matrix ${\cal M}$ may be 
approximately block diagonalized into a light, mostly left-handed
block, $M_\nu$, and a heavy, mostly right-handed block, $M_R$, 
\begin{eqnarray}
	\left(\begin{array}{cc}
		M_\nu & 0 \\
		0 & M_R \\
		\end{array}\right) 
	&\simeq & \left(\begin{array}{cc}
		1 & \xi^\dagger \\
		-\xi & 1\\
		\end{array}\right)^T
	\left(\begin{array}{cc}
			M_{LL}^\dagger & M_{LR}\\
			M_{LR}^T & M_{RR}\\
			\end{array}\right)
	\left(\begin{array}{cc}
		1 & \xi^\dagger \\
		-\xi & 1\\
		\end{array}\right) \label{eq:block1}\\
	&& \nonumber \\
	&\simeq &
	\left(\begin{array}{cc}
		M_{LL}^\dagger-
		M_{LR}M_{RR}^{-1}M_{LR}^T & 0 \\
		0 & M_{RR} \\
		\end{array}\right) \label{eq:block2}
\end{eqnarray}
where $\xi = M_{RR}^{-1}M_{LR}^T$.  The corrections to 
Eqs.~(\ref{eq:block1}) and (\ref{eq:block2})
are suppressed by $\xi$ and may be neglected if
$\left|\xi_{ij}\right|\ll 1$.  (In our numerical work below, the 
magnitudes of the elements of $\xi$ are of order $10^{-5}$ or smaller,
so the approximation is well-justified.)
The above result for $M_\nu$ was quoted in Eq.~(\ref{eq:mnu1})
and is the general expression for the Type II seesaw mechanism.
While the Dirac mass matrix $M_{LR}$ is of the same order 
of magnitude as its counterpart for the charged leptons (``$M_\ell$''), 
its contribution to $M_\nu$ is suppressed 
by $\xi$.

The six physical neutrino states obtained by diagonalizing ${\cal M}$ are
all Majorana neutrinos~\cite{kayser}, a fact that is 
responsible for the symmetry of 
${\cal M}$ about the diagonal.  The two $3\times 3$ blocks in 
Eq.~(\ref{eq:block2}) are also symmetric and may each be diagonalized 
with a unitary matrix, yielding
\begin{eqnarray}
	M_\nu^\textrm{\scriptsize diag} &=& 
		V_L^{\nu \dagger} M_\nu V_L^{\nu *}  \; ,
	\label{eq:mnudiag}\\
	M_R^\textrm{\scriptsize diag} &=&
		V_R^{\nu T}M_{RR}V_R^\nu \; ,
	\label{eq:mrdiag}
\end{eqnarray}
where the elements of the diagonal matrices are real and positive.

The unitary matrices $V_{L,R}^\ell$ and $V_{L,R}^\nu$
used to diagonalize the charged and neutral lepton mass matrices
may be combined to give the MNS matrices,
\begin{eqnarray}
	\widetilde{\cal U}_L^{MNS} &=& V_L^{\ell\dagger}V_L^\nu \; ,
		\label{eq:MNStilde1}\\
	\widetilde{\cal U}_R^{MNS} &=& V_R^{\ell\dagger}V_R^\nu \; ,
		\label{eq:MNStilde2}
\end{eqnarray}
where the tildes indicate that the matrices may still be
``rephased'' to bring them into the conventional form.
The rephasing procedure for the MNS matrices is accomplished by
multiplying the expressions in 
Eqs.~(\ref{eq:MNStilde1}) and (\ref{eq:MNStilde2})
on the left and right by diagonal phase matrices,
\begin{eqnarray}
	{\cal U}_L^{MNS} &=& B^\dagger \widetilde{\cal U}_L^{MNS}
		S_L \; , 
		\label{eq:ULMNS_rephased} \\
	{\cal U}_R^{MNS} &=& B^\dagger \widetilde{\cal U}_R^{MNS}
		S_R \; ,
		\label{eq:URMNS_rephased}
\end{eqnarray}
where
\begin{eqnarray}
	B= \left(\begin{array}{ccc}
		e^{i\rho_1} & 0 & 0 \\
		0 & e^{i\rho_2} & 0 \\
		0 & 0 & e^{i\rho_3} \\
		\end{array}\right)\; , ~~~
	S_L= \left(\begin{array}{ccc}
		e^{im_1 \pi} & 0 & 0 \\
		0 & e^{im_2 \pi} & 0 \\
		0 & 0 & 1 \\
		\end{array}\right)\; , ~~~
	S_R= \left(\begin{array}{ccc}
		e^{im_1^R \pi} & 0 & 0 \\
		0 & e^{im_2^R \pi} & 0 \\
		0 & 0 & e^{im_3^R \pi}  \\
		\end{array}\right) ,
\end{eqnarray}
with $m_i$ and $m_i^R$ integers.
The above rephasing procedure differs from its counterpart in the quark sector
in two respects.  In the first place, since the neutrinos
are Majorana particles, the ``phase'' matrices $S_L$ and $S_R$ are
actually only ``sign'' matrices, with factors of
$\pm 1$ appearing along the diagonals.
A second difference compared to the quark case is that 
$S_L$ and $S_R$ are distinct matrices, whereas in the
quark case the analogous matrices are equal~\cite{kiersLR}.
We may use the above results to write the charged current couplings
in the Lagrangian in terms of the physical mass eigenstates.  
Ignoring
terms further suppressed by $\xi$ in Eq.~(\ref{eq:block2}), we have~\cite{kayser}
\begin{eqnarray}
	{\cal L}_{CC} \simeq -\frac{g}{\sqrt{2}}\overline{e}_L{\cal U}_L^{MNS}
		\gamma_\mu\nu_L W_L^{\mu -}
		-\frac{g}{\sqrt{2}}\overline{e}_R{\cal U}_R^{MNS}
		\gamma_\mu\nu_R W_R^{\mu -}+ \textrm{h.c.},
\end{eqnarray}
where
\begin{eqnarray}
	\nu_{L,R} &=& S_{L,R}^\dagger V_{L,R}^{\nu\dagger}\nu_{L,R}^\prime
		\; , \\
	e_{L,R} &=& B^\dagger V_{L,R}^{\ell\dagger}e_{L,R}^\prime \; .
\end{eqnarray}

It is in general possible to parametrize the rephased 
left-handed MNS matrix in terms of three non-removable CP-odd phases.
One of these phases is analogous to the usual CKM phase in the 
left-handed CKM matrix.  The other two phases are novel, compared
to the quark sector, and their presence is due to the Majorana nature
of the neutrinos.  Attempts to remove these Majorana phases result
in their appearing elsewhere in the theory (in the diagonalized
neutrino masses, for example).  A useful parameterization of the left-handed
MNS matrix is~\cite{kayser}
\begin{eqnarray}
	{\cal U}_L^{MNS} = {\cal U}^{(0)}(\theta_{12},\theta_{23},
		\theta_{13},\delta_L) A_L,
		\label{eq:ULMNS}
\end{eqnarray}
where $A_L=\textrm{diag}(e^{i\alpha_1/2},e^{i\alpha_2/2},1)$ and
\begin{eqnarray}
	{\cal U}^{(0)}(\theta_{12},\theta_{23},\theta_{13},\delta_L) & & 
		\nonumber \\
		&\!\!\!\!\!\!\!\! = & \left(\begin{array}{ccc}
		c_{12}c_{13} & s_{12}c_{13} & s_{13}e^{-i\delta_L} \\
		-s_{12}c_{23}-c_{12}s_{23}s_{13}e^{i\delta_L} &
			c_{12}c_{23}-s_{12}s_{23}s_{13}e^{i\delta_L} &
			s_{23}c_{13} \\
		s_{12}s_{23}-c_{12}c_{23}s_{13}e^{i\delta_L} &
			-c_{12}s_{23}-s_{12}c_{23}s_{13}e^{i\delta_L} &
			c_{23}c_{13} \\
		\end{array}\right) \! .
	\label{eq:U_zero_mns}
\end{eqnarray}
The phase $\delta_L$ in the above expression is analogous to the usual
CP-odd ``Dirac''
phase in the quark sector, while the phases $\alpha_1$ and $\alpha_2$
are Majorana phases.  The right-handed MNS matrix contains six phases
in general, three of which may be taken to be Majorana phases.  A
convenient parameterization is as follows,
\begin{eqnarray}
	{\cal U}_R^{MNS} = A_\ell^\dagger \;
		{\cal U}^{(0)}(\theta_{12}^R,\theta_{23}^R,
		\theta_{13}^R,\delta_R) A_R,
		\label{eq:RHMNS}
\end{eqnarray}
where $A_R=\textrm{diag}(e^{i\alpha_1^R/2},e^{i\alpha_2^R/2},
e^{i\alpha_3^R/2})$ and 
$A_\ell = \textrm{diag}(e^{i\zeta_1},e^{i\zeta_2},1)$.

The left-handed MNS matrix has been probed through neutrino oscillation experiments,
which have placed relatively tight constraints on the three mixing angles $\theta_{ij}$.
The left-handed phases $\delta_L$ and $\alpha_{1,2}$ have not as yet been constrained by
experiment.  Neutrinoless double beta decay experiments could well
be used to probe combinations of the left-handed phases (depending on the 
ordering of the light neutrino
masses -- i.e., ``normal'' or ``inverted'' -- and
on the magnitudes of the masses).
In fact, such experiments
play a central role in neutrino physics, 
since the decays in question can only proceed if neutrinos are Majorana (as opposed to Dirac) particles.
The amplitudes for such decays are proportional to $m_{\beta\beta}$, the effective neutrino mass
for neutrinoless double beta decay~\cite{kayser},
\begin{eqnarray}
	m_{\beta\beta}&=&\left|\sum_j \left({\cal U}_{L_{1j}}^{MNS}\right)^2 m_j\right| \nonumber \\
		&=& \left|m_1c_{12}^2c_{13}^2 + m_2s_{12}^2c_{13}^2e^{i(\alpha_2-\alpha_1)}
			+m_3s_{13}^2e^{-i(\alpha_1+2\delta_L)}\right| \; .
	\label{eq:eff_nu_mass} 
\end{eqnarray}
Future experiments could probe $m_{\beta\beta}$ at the ${\cal O}(10^{-2}~\mbox{eV})$
level (see, for example, Ref.~\cite{Bilenky:2002aw}, as well
as~\cite{neutrino_unbound}).\footnote{There is 
controversial evidence of a non-zero neutrinoless double beta 
decay signal with $m_{\beta\beta}$ of 
order $0.5$~eV~\cite{Klapdor-Kleingrothaus:2001ke,Klapdor-Kleingrothaus:2004ge}.  
See also Refs.~\cite{Feruglio:2002af,Aalseth:2002dt}.}
Evidently $m_{\beta\beta}$ could be a sensitive probe of the MNS phases
if the mixing angles and masses were well known.\footnote{In
our notation the expression for $m_{\beta\beta}$ contains the phase combinations
$\alpha_2-\alpha_1$ and $\alpha_1+2\delta_L$.  It is possible to rephase the MNS matrix in such a way
that the $1$-$3$ element of ${\cal U}^{(0)}$ is real and the two Majorana phases in $A_L$ occur in the $2$-$2$
and $3$-$3$ elements~\cite{Choubey:2005rq}.  In that notation $m_{\beta\beta}$ depends only on the (two)
Majorana phases contained in $A_L$.  We have verified that the relations between the phases
used in the two approaches are such that one obtains the same physical value 
for $m_{\beta\beta}$ in either approach.}  In the numerical work below we will calculate $m_{\beta\beta}$
for this model to determine prospects for future experiments.

\subsection{Simplification of the Yukawa Couplings}
\label{sec:simp_Yuk}

It is often possible to simplify the Yukawa couplings in a model
through unitary transformations that leave physical quantities, such as
masses and mixings, unchanged.  In the present case, 
unitary transformation on the
Yukawa coupling matrices $G$, $H$ and $F$ may be used to
reduce the number of parameters required to specify the model.
As noted above, in order to satisfy the parity symmetry in 
Eq.~(\ref{eq:parity}), $G$ and $H$ must both be Hermitian.  Furthermore,
$F$ may be taken to be (complex)
symmetric.  For three generations of leptons, this
means that 30 real parameters are required to specify the elements
in the Yukawa matrices.  In principle, several of these degrees of freedom
are spurious and may be ``rotated away'' by an appropriate unitary rotation.
To see this, note that the diagonalized mass matrices and MNS matrices
are invariant under the rotations
\begin{eqnarray}
	F &\to & X^T F X , \nonumber \\
	G &\to & X^\dagger G X , \label{eq:Xtrans}\\
	H &\to & X^\dagger H X , \nonumber
\end{eqnarray}
where $X$ is a $3\times 3$ unitary matrix.  Furthermore,
these rotations preserve the
essential symmetries of the matrices, leaving $G$ and $H$ Hermitian
and $F$ symmetric.  In principle, one could use $X$ to diagonalize
one of the three Yukawa matrices, significantly decreasing the number
of parameters required to specify the model.
Within the context of the horizontal symmetry scheme that we employ,
however, the above transformations affect the scaling of the various 
terms.  For this reason we do not diagonalize any of the Yukawa matrices,
choosing instead to use a phase rotation in Eq.~(\ref{eq:Xtrans}) to 
remove one phase from $F$ ($\arg(F_{22}$)) and two from $H$
($\arg(H_{12})$ and $\arg(H_{13})$).  This reduces the
number of parameters required to specify the Yukawa couplings to 27.
In the numerical work below, a Monte Carlo algorithm is used
to search the 27-dimensional parameter space
to determine sets of parameters that are consistent with experimental
constraints on the lepton masses and mixings.

\subsection{Model with a broken $U(1)$ symmetry}
\label{sec:khasanov_perez}

One attractive way to account for the observed hierarchies in the quark
and lepton Yukawa couplings is to attribute them to a broken horizontal
symmetry~\cite{froggatt,nir1993}.  In models with a broken horizontal
symmetry, the various Yukawa couplings
are suppressed by powers of one or more small parameters, where the powers
are determined by the charges of the relevant fields under the horizontal
symmetry group.
Khasanov and Perez~\cite{khasanov} recently formulated a model that
uses a broken horizontal $U(1)$ symmetry to address two known problems
that occur in the LRM if one attempts to take $v_R$ to be only moderately large
(of order 20~TeV, say).  
The two problems are associated with the two terms appearing
in Eq.~(\ref{eq:mnu1}) -- as noted in the Introduction,
both terms run into trouble for moderate values of $v_R$
unless they are suppressed in some manner.
The first term in this expression is particularly troublesome -- although it
is somewhat suppressed due to the VEV seesaw, it is still far too large.
For $v_R$ of order 20~TeV, minimization
of the Higgs potential yields $v_L\sim k^2/v_R\sim(246.2~\mbox{GeV})^2/(20~\mbox{TeV})\sim
{\cal O}(1~\mbox{GeV})$, assuming all dimensionless coefficients in the 
Higgs potential to be of order unity (see Ref.~\cite{kiers_HiggsLR},
for example).  
If the Yukawa matrix $F$ in (\ref{eq:MLRetc})
is of order unity, then $M_{LL}$ will be of order 1~GeV,
approximately nine or ten orders of magnitude larger than the neutrino mass scale.
The second term in Eq.~(\ref{eq:mnu1}) is also too large if
$v_R$ is of order 20~TeV.  Assuming $F$ to be of order unity and the
largest elements of $M_{LR}$ to be of order $m_\tau$, we find
that ``$-M_{LR}M_{RR}^{-1}M_{LR}^T$'' generically has elements of order 
$m_\tau^2/v_R\sim(1.777~\mbox{GeV})^2/(20~\mbox{TeV})\sim {\cal O}(0.1~\mbox{MeV})$, 
which are still too
large from a phenomenological point of view.\footnote{One could improve the situation
by assuming that the Yukawa matrix $G\sim 0$.  In that case, the largest elements in $H$
are of order $m_\tau/k_1$ and we have $M_{LR}\sim H k_2\sim m_\tau\times (k_2/k_1)$.
As noted above, it is natural to assume $k_2/k_1\sim m_b/m_t$~\cite{kiersLR}, in which case
the largest elements in ``$-M_{LR}M_{RR}^{-1}M_{LR}^T$'' are of order 10's of eV
for $v_R=20$~TeV.  In the horizontal symmetry scheme that we employ below,
$G$ is in fact suppressed relative to $H$, leading to a similar result 
(see Eq.~(\ref{eq:mnu_order_of_mag}) and the discussion that follows).}

A model with a 
broken horizontal $U(1)$ symmetry offers a solution to both of the problems
noted above. 
At high energies the model contains a new scalar $S$ as well as several new
heavy fermions.  Most of the Yukawa terms in Eq.~(\ref{eq:yuk}) are not
present in the high energy theory because they
do not respect the $U(1)$ symmetry.  Instead, such terms
descend from nonrenormalizable
terms in the low energy effective
theory obtained by integrating out the heavy fermions.
As a result of this procedure, the Yukawa terms contain various
powers of a small symmetry breaking parameter $\epsilon=\langle S\rangle/M$, 
where $M$ is the mass scale of the heavy fermions. 
The power of $\epsilon$ for a given term in the Lagrangian
is determined by the $U(1)$ charges of the fields coupled together in that
term.  The Yukawa couplings scale as follows,
\begin{eqnarray}
	F_{ij} &=& \widetilde{F}_{ij}
		\epsilon^{|Q(\Delta_L)+Q(L_L^i)+Q(L_L^j)|} , \nonumber \\
	G_{ij} &=& \widetilde{G}_{ij}
		\epsilon^{|Q(L_R^j)-Q(L_L^i)+Q(\phi)|} , 	\label{eq:Yuk_scale} \\
	H_{ij} &=& \widetilde{H}_{ij}
		\epsilon^{|Q(L_R^j)-Q(L_L^i)-Q(\phi)|} , \nonumber
\end{eqnarray}
where the quantities with the tildes are taken to be of order unity in 
magnitude.  In our numerical work we adopt the following charge assignments (see also
Refs.~\cite{kiers_HiggsLR,khasanov}),
\begin{eqnarray}
	Q(\Delta_L)&=&-Q(\Delta_R)=-8 , \nonumber \\
	Q(\phi)&=& -2 ,	\label{eq:charges} \\
	Q(L_L^{1,2,3})&=&-Q(L_R^{1,2,3}) = 6, 4, 3 , \nonumber
\end{eqnarray}
yielding
\begin{eqnarray}
	F\sim \left(\begin{array}{ccc}
		\epsilon^4 & \epsilon^2 & \epsilon \\
		\epsilon^2 & 1 & \epsilon \\
		\epsilon & \epsilon & \epsilon^2 \\
		 \end{array}\right) ,~~~
	G\sim \left(\begin{array}{ccc}
		\epsilon^{14} & \epsilon^{12} & \epsilon^{11} \\
		\epsilon^{12} & \epsilon^{10} & \epsilon^9 \\
		\epsilon^{11} & \epsilon^9 & \epsilon^8 \\
		 \end{array}\right) ,~~~
	H\sim \left(\begin{array}{ccc}
		\epsilon^{10} & \epsilon^{8} & \epsilon^{7} \\
		\epsilon^{8} & \epsilon^{6} & \epsilon^5 \\
		\epsilon^{7} & \epsilon^5 & \epsilon^4 \\
		 \end{array}\right) ,
	\label{eq:FGH_eps}
\end{eqnarray}
where coefficients of order unity have been omitted.
For the purpose of our numerical work we set $\epsilon=0.3$, 
as in Ref.~\cite{kiers_HiggsLR}, a value that automatically 
gives charged lepton masses in the correct range.
The Higgs potential of the low energy effective 
theory also contains
terms that break the $U(1)$ symmetry, leading to a suppression
of many of the dimensionless coefficients in the Higgs potential.
Reference~\cite{kiers_HiggsLR} contains a thorough discussion
of the Higgs sector of the LRM with a broken horizontal symmetry.\footnote{In 
that paper it was shown that a phenomenologically acceptable
Higgs spectrum emerges if explicit CP violation is
allowed in the Higgs potential.  This is to be contrasted with
the case in which the Higgs potential is CP-invariant.  In that case,
non-negligible CP violation in the vacuum state is generically accompanied by
non-SM-like neutral Higgs bosons at the weak scale
with flavour non-diagonal couplings~\cite{bgnr}.}
With the charge assignments noted in (\ref{eq:charges}), minimization
of the Higgs potential leads to the following expression for $v_L$~\cite{kiers_HiggsLR}, 
\begin{eqnarray}
	v_L = \gamma \epsilon^{20}k_1^2/v_R ,
		\label{eq:magvL}
\end{eqnarray}
where $\gamma$ depends on various dimensionless coefficients in the Higgs potential
and is generically of order unity.  For $\gamma$ of order unity (and setting
$\epsilon=0.3$), we have
\begin{eqnarray}
	v_L \sim \frac{0.1~\textrm{eV}}{v_R/(20~\textrm{TeV})} ,
		\label{eq:vLapprox}
\end{eqnarray}
which is phenomenologically viable.  Thus the $U(1)$ model
successfully deals with the first of the two problems noted above.

The $U(1)$ model also deals successfully with the fact
that the second term in Eq.~(\ref{eq:mnu1}) is generically too large.
Before considering this term, let us examine the charged
lepton mass matrix, $M_{\ell}$, given in Eq.~(\ref{eq:mlep}).
To a good approximation, one may
neglect the contribution of $G$ to $M_{\ell}$, since this contribution is
suppressed by a factor of approximately $\epsilon^4\times (k_2/k_1)\sim {\cal O}(10^{-4})$ 
relative to that of $H$.
The situation is different for the neutrino Dirac mass matrix 
$M_{LR}$, since 
the roles of $k_1$ and $k_2$ are essentially reversed in this case.
In fact,  $G$ and $H$ contribute comparable amounts to $M_{LR}$
(since $k_2/k_1\sim {\cal O}(\epsilon^4)$) and we find that
$M_{LR}= \left(Gk_1+Hk_2e^{-i\alpha}\right)/\sqrt{2} \sim M_\ell \times \epsilon^4$ as
an order of magnitude estimate.
Noting that
\begin{eqnarray}
	F^{-1} \sim \left(\begin{array}{ccc}
		1 & 1 & 1/\epsilon \\
		1 & 1 & \epsilon \\
		1/\epsilon & \epsilon & \epsilon^2 \\
		 \end{array}\right) 
	\label{eq:finv}
\end{eqnarray}
(where a ``1'' denotes an element of order unity), we have
\begin{eqnarray}
	M_\nu = M_{LL}^\dagger-
		M_{LR}M_{RR}^{-1}M_{LR}^T 
		&\sim & \frac{k_1^2}{\sqrt{2} v_R}\left(\begin{array}{ccc}
		\epsilon^{24} & \epsilon^{22} & \epsilon^{21} \\
		\epsilon^{22} & \epsilon^{20} & \epsilon^{21} \\
		\epsilon^{21} & \epsilon^{21} & \epsilon^{22} \\
		 \end{array}\right) 
		+\frac{k_1^2}{\sqrt{2}v_R}\left(\begin{array}{ccc}
		\epsilon^{24} & \epsilon^{22} & \epsilon^{21} \\
		\epsilon^{22} & \epsilon^{20} & \epsilon^{19} \\
		\epsilon^{21} & \epsilon^{19} & \epsilon^{18} \\
		 \end{array}\right)\\
		&\sim & \frac{k_1^2}{\sqrt{2}v_R}\left(\begin{array}{ccc}
		\epsilon^{24} & \epsilon^{22} & \epsilon^{21} \\
		\epsilon^{22} & \epsilon^{20} & \epsilon^{19} \\
		\epsilon^{21} & \epsilon^{19} & \epsilon^{18} \\
		 \end{array}\right) 
		\label{eq:mnu_epsilon}\\
		&\sim & \frac{1}{v_R/(20~\textrm{TeV})}\left(\begin{array}{ccc}
		0.0006 & 0.007 & 0.02 \\
		0.007 & 0.07 & 0.2 \\
		0.02 & 0.2 & 0.8 \\
		 \end{array}\right) ~\textrm{eV} ,
		\label{eq:mnu_order_of_mag}
\end{eqnarray}
where the numerical values in the last line should be understood as being
very approximate.\footnote{Given the large exponents in this expression, 
one might worry about the sensitivity of these results
to small deviations in the parameter $\epsilon$.  While it is true that a small change in $\epsilon$ would 
produce a larger effect in $M_\nu$, the effect on $M_\nu$
could be partly compensated by adjusting $v_R$.}

The two terms in the above expression for $M_\nu$ contribute at
approximately the same level and combine to yield neutrino masses
that are of the correct order of magnitude.
It is interesting to see how the horizontal symmetry model 
deals with the fact
that the largest elements in the Type I seesaw part of 
$M_\nu$ (the second term) are generically of order $m_\tau^2/v_R\sim$~0.1~MeV.  
The main suppression of such elements in the $U(1)$ model follows from the
fact that the largest terms in $M_{LR}$ are now of order $\epsilon^4 m_\tau$,
instead of $m_\tau$,
as noted in the discussion above Eq.~(\ref{eq:finv}).  
A further suppression is due to the particular structures of $F^{-1}$ and $M_{LR}$.\footnote{For
example, one contribution to the $3$-$3$ element in $M_\nu$ comes from
the $3$-$3$ elements of $M_{LR}$ and $F^{-1}$.  While
$M_{LR,33}$ is generically the largest element of $M_{LR}$, 
$F^{-1}_{33}\sim \epsilon^2$, so the combined contribution is of
order $\epsilon^2\times(\epsilon^4 m_\tau)^2/v_R$.}

The LRM with a broken $U(1)$ symmetry is thus able to reproduce the gross features
of the lepton mass spectra, 
yielding the correct orders of magnitude for the charged lepton
masses as
well as an appropriate mass scale for the light neutrinos.
In the following, we consider whether the model is able to accommodate
the experimental values for the light neutrino mass-squared
differences and mixing angles.  In fact, there is potentially
a difficulty in this regard, as was pointed out by
Khasanov and Perez~\cite{khasanov} -- the
$1$-$2$ element in Eq.~(\ref{eq:mnu_epsilon}) is generically suppressed
relative to the $2$-$2$ element,
indicating a possible difficulty in obtaining
a large $1$-$2$ mixing angle.
Nevertheless, we shall show that it is in fact possible to satisfy the experimental
constraints on all three mixing angles
and on the mass-squared differences
in this model.  We offer some further comments on this issue in Appendix~\ref{sec:appendix2}.
As is noted there, the numerical procedure favours
neutrino mass matrices that have a quasi-degenerate
$2$-$3$ block, with some or all elements in the block suppressed relative
to Eq.~(\ref{eq:mnu_order_of_mag}).

\section{Numerical Study}
\label{sec:numerical}

In this section we perform a numerical analysis
of the model.  The goal of the numerical work is to find sets
of values for the various Higgs VEVs and Yukawa couplings such
that the experimental constraints on the lepton masses and mixings are 
satisfied.  Equation~(\ref{eq:Yuk_scale}) expresses the three Yukawa matrices
$F$, $G$ and $H$ in terms of rescaled (order unity) Yukawa 
couplings ($\widetilde{F}_{ij}$, etc.) multiplied by appropriate powers of $\epsilon$.  
Many of the Yukawa couplings are complex.  Recalling that $F$ is complex-symmetric
and that $G$ and $H$ are both Hermitian, we define phases as follows,
\begin{eqnarray}
	\widetilde{F}_{ij} &=& \widetilde{F}_{ji} = \left|\widetilde{F}_{ij}\right| e^{i\theta^F_{ij}} 
		~~~~~~~(j\geq i), \nonumber \\
	\widetilde{G}_{ij} &=& \widetilde{G}^*_{ji} = \left|\widetilde{G}_{ij}\right| e^{i\theta^G_{ij}} 
		~~~~~~~(j>i), \label{eq:FGH_phases} \\
	\widetilde{H}_{ij} &=& \widetilde{H}^*_{ji} = \left|\widetilde{H}_{ij}\right| e^{i\theta^H_{ij}} 
		~~~~~~~(j>i). \nonumber
\end{eqnarray}
The diagonal elements of $\widetilde{G}$ and $\widetilde{H}$ are real, but possibly negative.
As described in Sec.~\ref{sec:simp_Yuk}, 
unitary rotations may be used to simplify the Yukawa matrices
without affecting the lepton masses or the MNS matrices.
We use such rotations to eliminate one phase in $F$ and two in $H$, setting
$\theta^F_{22}=\theta^H_{12}=\theta^H_{13}=0$.
Thus, there are a total of 27 parameters used to describe the 
Yukawa matrices, nine of which are phases.  In our numerical work, we allow
the magnitudes of the scaled Yukawa couplings to be 
in the range zero to three and the phases to be in the range zero to $2\pi$.

For the Higgs VEVs, we use Eq.~(\ref{eq:kkpr1}) to fix the sum
$k_1^2+k_2^2$ and take the ratio $k_2/k_1$ to be $3/181$ 
(as in Refs.~\cite{kiersLR,kiers_HiggsLR}). 
We consider two cases for $v_R$, taking $v_R=20$~TeV and $v_R=50$~TeV.
$v_L$ is defined through Eq.~(\ref{eq:magvL}), where
we take $\gamma$ to be chosen randomly
in the range zero to two.
It remains to consider the phases of the Higgs VEVs, $\alpha$ and $\theta_L$ (see Eq.~(\ref{eq:vevs})).
Correlations between $\alpha$, $\theta_L$ and $v_L$ were studied in Ref.~\cite{kiers_HiggsLR}.
Since the observed correlations were not very strong, 
we simply allow $\alpha$ and $\theta_L$ to take any values in the range zero to $2\pi$.
Adding $\alpha$, $\theta_L$ and $v_L$ to the 27 Yukawa coupling parameters, we find that we
have a total of 30 parameters to fix.  This number exceeds the number of experimental constraints
on the model, which come from the charged lepton masses (3), the neutrino mass-squared differences (2)
and the neutrino mixing angles (3).\footnote{We do not include the LSND results in our analysis.}
Clearly it will not be possible to fix the 30 ``input parameters'' uniquely.  Nevertheless, 
a Monte Carlo approach can be used to find sets of input parameters that yield masses and mixings consistent with
experiment.  Once these experimental constraints have been satisfied, other quantities in the model --
such as left-handed neutrino phases and right-handed neutrino masses and mixings -- can be calculated.

\subsection{Monte Carlo Algorithm}

The Monte Carlo approach that we use is similar to that described in Ref.~\cite{kiersLR}.
A rough summary of the procedure is as follows.
Sets of input parameters are chosen randomly and then used to form the
various mass matrices.  Diagonalization of these mass matrices yields 
theoretical values for the masses and mixing angles, which are then
compared with their experimental counterparts.
Specifically, we calculate the charged lepton masses, the neutrino mass-squared differences
and the squares of the sines of the mixing angles 
and compare these to the experimental values 
described in Table~\ref{tab:masses_mixings}.\footnote{For the charged leptons
we adopt relative uncertainties of
$5\times 10^{-4}$, which are larger than the experimental uncertainties~\cite{PDG2004}.
This is done for the sake of the efficiency of our Monte Carlo algorithm.}
A quantitative measure of the ``goodness of fit'' is provided
by the quantity $\chi^2$,
\begin{eqnarray}	
	\chi^2 = \sum_{i=1}^8 \frac{(y_i^\textrm{\scriptsize{exp}}-y_i)^2}{\sigma_i^2} \; ,
\end{eqnarray}
where the sum runs over the five experimental constraints
$y_i^\textrm{\scriptsize{exp}}\pm \sigma_i$ in Table~\ref{tab:masses_mixings},
as well as three constraints coming from the charged lepton masses.  The associated values obtained numerically
are denoted $y_i$.
The Monte Carlo algorithm essentially hunts around
the parameter space seeking to reduce $\chi^2$ to an acceptable value.  A set of input parameters is declared
to be a solution if $\left| y_i^\textrm{\scriptsize{exp}}-y_i\right| \leq \sigma_i$ for all $i$.

\begin{table*}
\caption{Experimental constraints for neutrino masses and mixings
used in the numerical work.  The values
adopted for $y_i^\textrm{\scriptsize{exp}}$ and $\sigma_i$
correspond to the $3\sigma$ ranges given in Table 1 of Ref.~\cite{Maltoni:2004ei}.  
We estimate the central values $y_i^\textrm{\scriptsize{exp}}$ by
bisecting the $3\sigma$ ranges.
Normal mass ordering is assumed in the numerical
work, so both of the mass-squared differences are taken to be positive.}
\begin{ruledtabular}
\begin{tabular}{cc}
	Quantity & $y_i^\textrm{\scriptsize{exp}}\pm\sigma_i$ \\ \colrule
	$\Delta m_{21}^2=m_2^2-m_1^2$ & $(8.15\pm 0.95)\times 10^{-5}$~eV$^2$ \\
	$\Delta m_{31}^2=m_3^2-m_1^2$ & $(2.35\pm0.95)\times 10^{-3}$~eV$^2$\\
	$\sin^2\theta_{12}$ & $0.305\pm0.075$\\
	$\sin^2\theta_{23}$ & $0.51\pm 0.17$\\
	$\sin^2\theta_{13}$ & $0.0235\pm 0.0235$\\
\end{tabular}
\end{ruledtabular}
\label{tab:masses_mixings}
\end{table*}

The relative uncertainties associated with the charged lepton masses are quite small.
Furthermore, the charged lepton mass matrix
only depends on $G$ and $H$ (see Eq.~(\ref{eq:mlep})).  These two factors make it
convenient to split the search algorithm into two phases, with the first phase
searching for Yukawa matrices $G$ and $H$ that yield acceptable charged lepton masses and the second phase
searching for a Yukawa matrix $F$ that results in acceptable neutrino masses and mixings.
Sometimes more than one acceptable matrix $F$ is found for a given pair of matrices $G$ and $H$.  
In such cases the sets of input parameters are considered to be
separate solutions, since in general they yield different neutrino mass matrices.

\begin{figure}[htb]
\resizebox{7.5in}{!}{\includegraphics*[1.2in,4.1in][8in,7in]{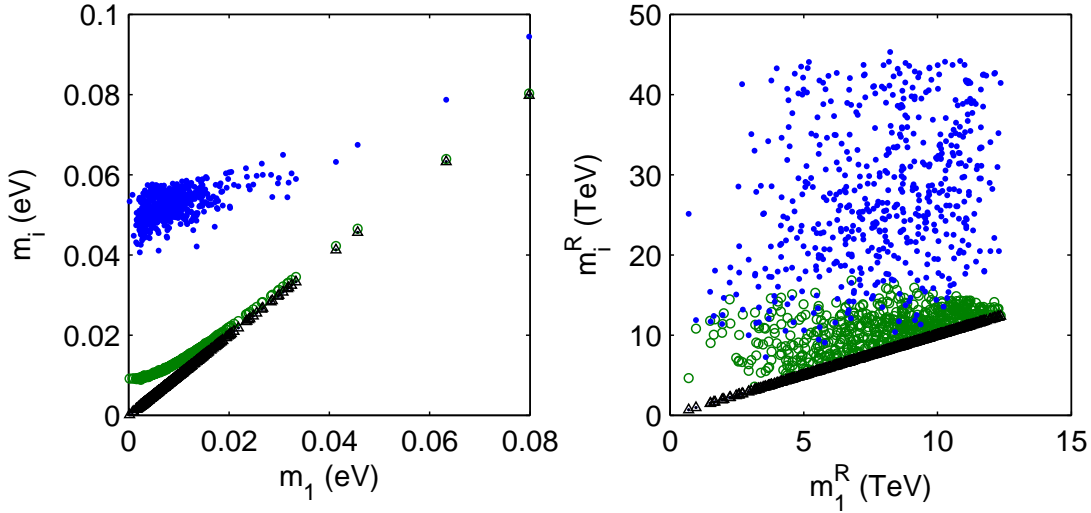}}
\caption{Light (left) and heavy (right) neutrino masses for $v_R=20$~TeV.
For each set of three masses, the lightest mass is indicated by a triangle, the intermediate one by
a circle and the heaviest by a dot.}
\label{fig:3masses}
\end{figure}
\begin{figure}[htb]
\resizebox{6in}{!}{\includegraphics{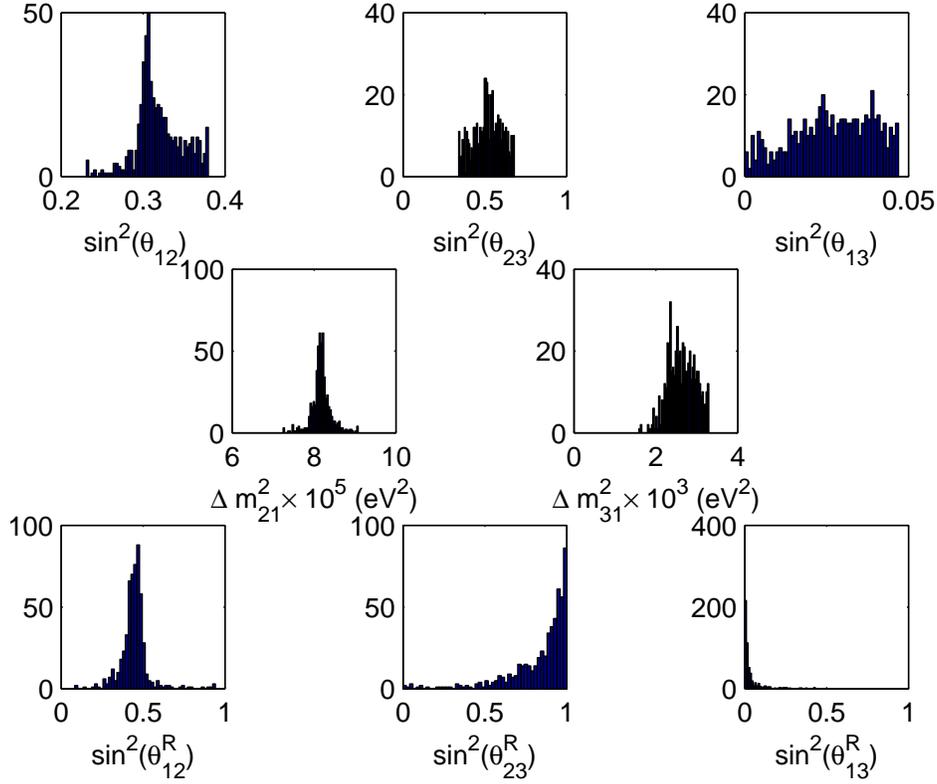}}
\caption{Frequency plots of mixing angles and mass-squared differences for $v_R=20$~TeV.
The top and bottom rows show results for the left- and right-handed mixing angles, respectively.
The middle row shows frequency plots for $\Delta m_{ij}^2=m_i^2-m_j^2$,
the mass-squared differences for the light neutrinos.
The plots in the top two rows satisfy the constraints noted in Table~\ref{tab:masses_mixings}.}
\label{fig:angles_Delm}
\end{figure}
\begin{figure}[t]
\resizebox{6in}{!}{\includegraphics{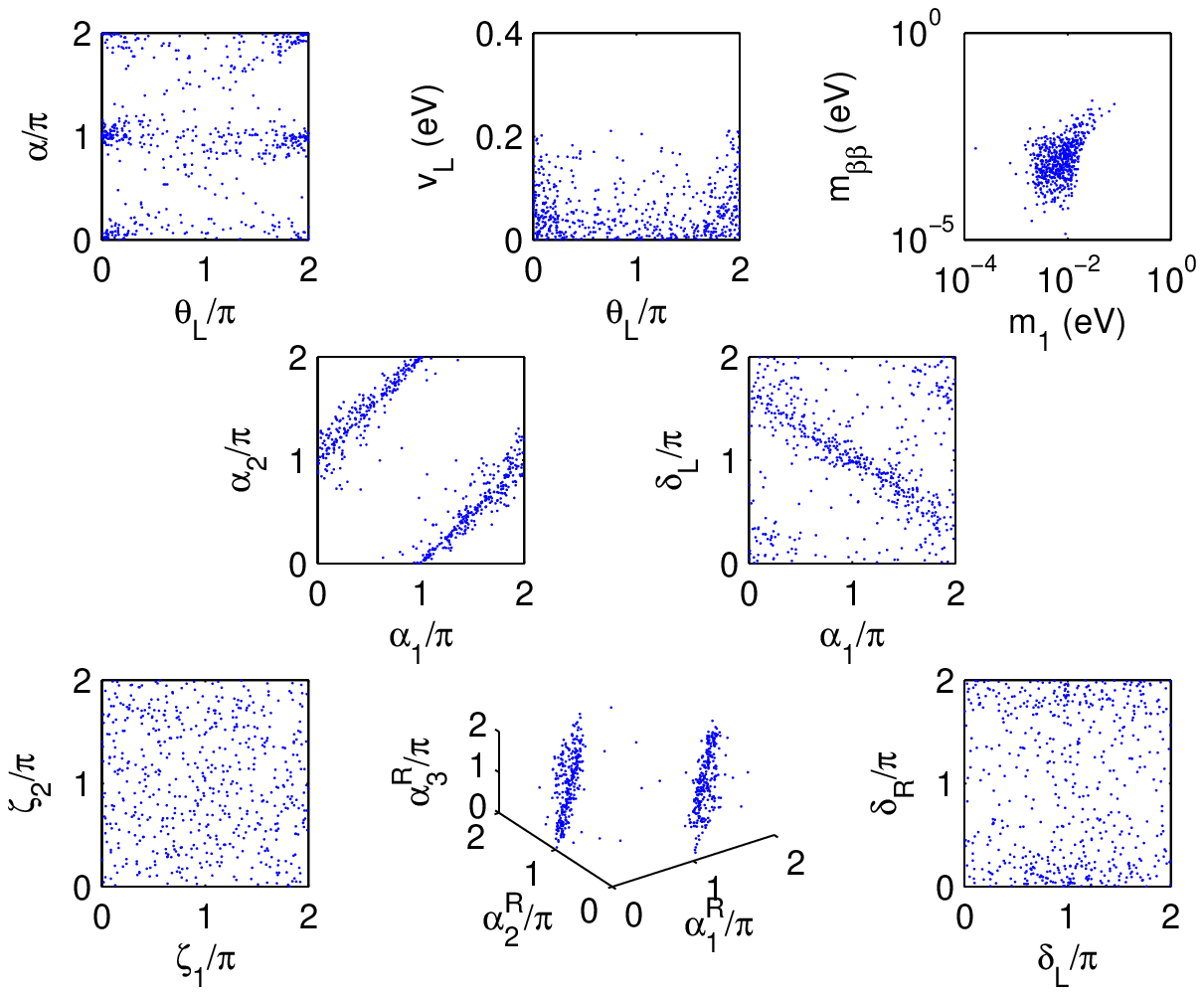}}
\caption{Plots of various phases, $v_L$ and $m_{\beta\beta}$ for $v_R=20$~TeV.
The effective neutrino mass $m_{\beta\beta}$ is defined in Eq.~(\ref{eq:eff_nu_mass}).}
\label{fig:phases}
\end{figure}
%

\subsection{Masses, mixings and phases for $v_R=20$~TeV and $v_R=50$~TeV}

In this subsection we summarize our results for neutrino masses, mixing angles and phases
for two choices for the right-handed scale, $v_R=20$~TeV and $v_R=50$~TeV.  
We also include some comments on $m_{\beta\beta}$, the
effective neutrino mass for neutrinoless double beta decay.  
In the following subsection we discuss some other
phenomenology of the model.

Figure~\ref{fig:3masses} shows the neutrino masses obtained for the case $v_R=20$~TeV.
The data were generated using the Monte Carlo algorithm outlined in the previous subsection.
Each particular set of ``input'' parameters (Yukawa couplings and Higgs VEVs) yields three light neutrinos
and three heavy neutrinos.
The plot on the left shows the light neutrino masses and indicates that
the model tends to favour
non-degenerate (as opposed to quasi-degenerate) light neutrinos.  
The plot on the right contains the results for the heavy neutrinos.
Approximate expressions
for the heavy neutrino masses are given in Appendix~\ref{sec:appendix1}.
As noted there, the two lightest right-handed neutrinos, $m_{1}^R$ and $m_{2}^R$, both 
have masses of order $\epsilon v_R$, while $m_3^R$ has a mass of order $v_R$.
This scaling is evident in Fig.~\ref{fig:3masses}.  Even though
the right-handed scale is $20$~TeV, it is not uncommon to have $m_1^R$
below $5$~TeV.
The mass splitting between $m_{1}^R$ and $m_{2}^R$ is typically
of order $\epsilon^2 v_R$, as is shown in Appendix~\ref{sec:appendix1}.

The plots in Fig.~\ref{fig:angles_Delm} show the mixing angles for the left- and right-handed MNS
matrices as well as the mass-squared differences for the light neutrinos.
The top two rows of plots show explicitly that the constraints on mass-squared differences and 
mixing angles in Table~\ref{tab:masses_mixings} are indeed satisfied by the model.  The
bottom row shows the right-handed mixing angles favoured by the model.  Appendix~\ref{sec:appendix1} contains
approximate expressions for each of the right-handed mixing angles, noting that
\begin{eqnarray}
	\sin^2\theta_{12}^R &\simeq & 0.5\left(1-{\cal O}(\epsilon)\right) \; , \nonumber \\
	\sin^2\theta_{23}^R &\simeq & 1-{\cal O}(\epsilon^2) \; ,
		\label{eq:app_sinsq_text} \\
	\sin^2\theta_{13}^R &\simeq & {\cal O}(\epsilon^4) \; . \nonumber
\end{eqnarray}
The interested reader is referred to this appendix for explicit expressions in terms of the relevant
Yukawa couplings.  The above expressions are consistent with the results indicated in Fig.~\ref{fig:angles_Delm}.

Figure~\ref{fig:phases} shows relations among the various phases appearing in the left- and right-handed MNS
matrices, as well as plots of $v_L$ and $m_{\beta\beta}$.  The upper left plot shows the
correlation between $\alpha$ and $\theta_L$ (the phases associated with the bidoublet and left-handed triplet
Higgs boson fields, respectively).  It is evident from the plot that the model (or at least the numerical procedure)
favours $\alpha$ near $0$ and $\pi$, although other values for $\alpha$ are not ruled out.
The middle plot in the top row shows the values obtained
for $v_L$.  As expected from Eq.~(\ref{eq:vLapprox}), 
$v_L$ is of order $0.1$~eV for $v_R=20$~TeV.  The upper right plot in the figure
shows that $m_{\beta\beta}$, the effective neutrino mass
for neutrinoless double beta decay (see Eq.~(\ref{eq:eff_nu_mass})), is typically of order
$0.001$ or $0.002$~eV for $v_R=20$~TeV in this model.\footnote{This plot may be compared with
Fig.~1 in Ref.~\cite{Aalseth:2002dt} (although slightly different experimental
ranges were used for that plot).}  Such values
are probably beyond the sensitivity of neutrinoless double beta decay experiments
of the near future.
To see why $m_{\beta\beta}$ is so small, consider its dependence on the
left-handed Majorana phases $\alpha_1$ and $\alpha_2$ and the ``Dirac'' phase
$\delta_L$ (shown in the middle pair of plots in Fig.~\ref{fig:phases}).
To a good approximation (i.e., taking $c_{13}^2\simeq 1$), Eq.~(\ref{eq:eff_nu_mass})
may be written as follows,
\begin{eqnarray}
	m_{\beta\beta} \simeq \left|m_3s_{13}^2+
		\left(m_1c_{12}^2-m_2s_{12}^2e^{i(\alpha_2-\alpha_1-\pi)}\right)
		e^{i(\alpha_1+2\delta_L)}\right| \; .
\end{eqnarray}
As is evident from Fig.~\ref{fig:phases}, to a good approximation,
$\alpha_2\approx \alpha_1+\pi~(\textrm{mod}~2\pi)$ and there is typically a partial cancellation
between the terms proportional to $m_1$ and $m_2$ in the above expression.  Also,
the $m_3$ term is suppressed by $\sin^2\theta_{13}$.
There could in principle be interference between
the terms, but the overall smallness of $m_{\beta\beta}$ would make
it difficult to use $m_{\beta\beta}$ as a probe of the phases involved.
The bottom row of plots in Fig.~\ref{fig:phases} shows the phases associated with the right-handed MNS matrix
(see Eq.~(\ref{eq:RHMNS})).  The right-handed Majorana phases $\alpha_1^R$
and $\alpha_2^R$ satisfy the approximate relation
$\alpha_2^R\approx \alpha_1^R+\pi~(\textrm{mod}~2\pi)$, a result that has been
derived analytically (see Appendix~\ref{sec:appendix1}).
Scatter plots for the other right-handed phases are also shown.

\begin{figure}[t]
\resizebox{7.5in}{!}{\includegraphics*[1.2in,4.1in][8in,7in]{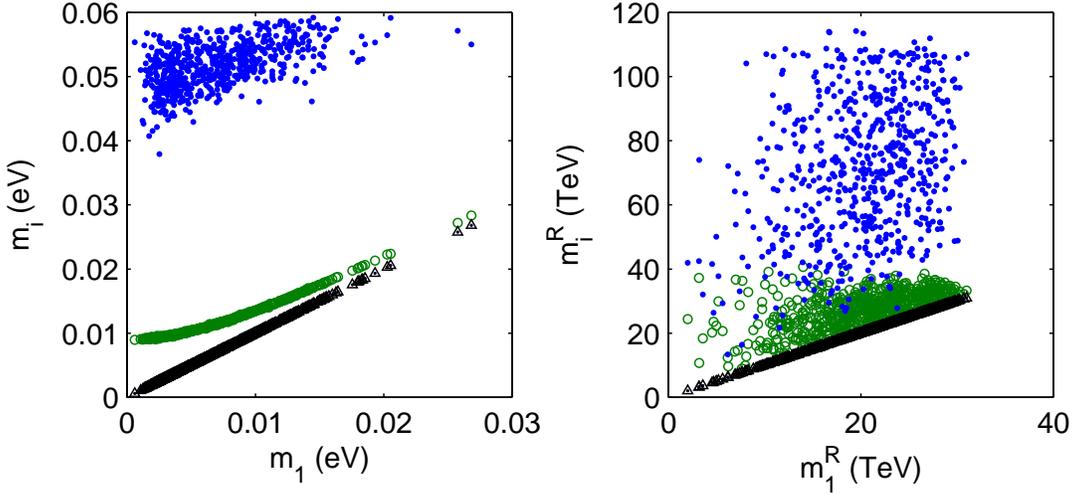}}
\caption{Same as in Fig.~\ref{fig:3masses}, but for $v_R=50$~TeV.  Note that the axes are different
here.}
\label{fig:3masses_50}
\end{figure}

A similar analysis has been performed for $v_R=50$~TeV and the results
are qualitatively similar to those for $v_R=20$~TeV.  The light and heavy
neutrino masses obtained for that case are shown in Fig.~\ref{fig:3masses_50}.
As might be expected, the heavy neutrino masses are larger than their
counterparts for $v_R=20$~TeV and the light neutrino masses are somewhat smaller, due to the
two seesaw mechanisms at work (see the discussion 
around Eq.~(\ref{eq:mnu_order_of_mag})).
Plots of the mixing angles and phases for $v_R=50$~TeV are similar to those in Figs.~\ref{fig:angles_Delm}
and \ref{fig:phases} and are not shown.  The values obtained
for $v_L$ and $m_{\beta\beta}$ are generally somewhat smaller for $v_R=50$~TeV than those
that were found for $v_R=20$~TeV.

%

\subsection{Some phenomenological implications}

In the previous subsection we discussed neutrinoless double beta decay in the context
of this model.
Let us now consider some other phenomenological consequences 
of this model, again taking $v_R$ to be in the range $20$ to $50$~TeV.
Incidentally, while we have not been much concerned 
in this work with the quark
sector of the theory, we should note that many authors have studied hadronic consequences for a right-handed scale
in the several-TeV range, such as effects on $B$-$\overline{B}$ and $K$-$\overline{K}$ mixing 
(see, for example, Refs.~\cite{bbs,kiersLR,ball1,ball2}, and references therein.)
We shall not consider such effects further here, but shall be mainly concerned with leptonic phenomenology.

First let us consider the effect that a ``moderate'' right-handed scale has on leptogenesis.
Leptogenesis provides a mechanism for generating the baryon asymmetry of the universe through
CP asymmetries involving leptons~\cite{Fukugita:1986hr,Luty:1992un,
Buchmuller:1996pa}.  Within the LRM, the asymmetries can occur in the decays of heavy (right-handed) neutrinos
to charged leptons and Higgs bosons, as well as in the decays of left-handed Higgs triplets to pairs of 
charged leptons (see Refs.~\cite{Hambye:2003ka,Antusch:2004xy,Chen:2004ww}).
The asymmetries arise through the interference of the tree-level diagrams 
with one-loop self-energy and vertex correction diagrams.
The asymmetries for the decay of the lightest right-handed neutrino may
be separated into Type I and II contributions
as follows~\cite{Hambye:2003ka,Antusch:2004xy,Chen:2004ww} (the reason for the ``Type I'' designation
for the first expression will be more apparent in a moment),
\begin{eqnarray}
        \epsilon_{\nu_{R_1}}^{I} &=& \frac{1}{4\pi(k_1^2+k_2^2)}\sum_{j\neq 1}
	       \frac{\textrm{Im}\left[\left(\widetilde{M}_{LR}^\dagger\widetilde{M}_{LR}\right)_{1j}^2\right]}
	            {\left(\widetilde{M}_{LR}^\dagger\widetilde{M}_{LR}\right)_{11}}
		    \sqrt{x_j}\left[1-\left(1+x_j\right)\ln\left(1+\frac{1}{x_j}\right)+\frac{1}{1-x_j}\right] \! , 
	       \label{eq:epsIexact}\\
	       & & \nonumber\\
	\epsilon_{\nu_{R_1}}^{II} &=& \frac{3 \, m_{1}^{R}}{4\pi(k_1^2+k_2^2)}
	       \frac{\textrm{Im}\left[\left(\widetilde{M}_{LR}^\dagger M_\nu^{II}\widetilde{M}_{LR}^*\right)_{11}\right]}
	            {\left(\widetilde{M}_{LR}^\dagger\widetilde{M}_{LR}\right)_{11}}
		    y\left[1-y\ln\left(1+\frac{1}{y}\right)\right] \! ,
               \label{eq:epsIIexact}
\end{eqnarray}
where $\widetilde{M}_{LR}\equiv M_{LR}V_R^\nu$ (with $V_R^\nu$ being the matrix that diagonalizes $M_{RR}$ --
see Eq.~(\ref{eq:mrdiag})).  Also, 
$x_j=(m_{j}^{R}/m_{1}^{R})^2$ and $y = (m_{\Delta_L}/m_{1}^{R})^2$, with $m_{\Delta_L}$ being the left-handed Higgs
triplet mass~(see Ref.~\cite{kiers_HiggsLR}).  It is instructive to consider the limits
$x_j\gg 1$ and $y\gg 1$,
in which case the asymmetries become
\begin{eqnarray}
        \epsilon_{\nu_{R_1}}^{I} &\simeq & \frac{3 \, m_{1}^{R}}{8\pi(k_1^2+k_2^2)}
	       \frac{\textrm{Im}\left[\left(\widetilde{M}_{LR}^\dagger M_\nu^{I}\widetilde{M}_{LR}^*\right)_{11}\right]}
	            {\left(\widetilde{M}_{LR}^\dagger\widetilde{M}_{LR}\right)_{11}} \; , 
	       \label{eq:epsIapprox}\\
		    & & \nonumber\\
	\epsilon_{\nu_{R_1}}^{II} &\simeq & \frac{3 \, m_{1}^{R}}{8\pi(k_1^2+k_2^2)}
	       \frac{\textrm{Im}\left[\left(\widetilde{M}_{LR}^\dagger M_\nu^{II}\widetilde{M}_{LR}^*\right)_{11}\right]}
	            {\left(\widetilde{M}_{LR}^\dagger\widetilde{M}_{LR}\right)_{11}}  \;  ,
	       \label{eq:epsIIapprox}
\end{eqnarray}
illustrating a nice symmetry between the two expressions~\cite{Antusch:2004xy}.  
(Recall that $M_\nu^{I}$ and $M_\nu^{II}$ are the Type I and II
contributions to the light neutrino mass matrix -- see Eq.~(\ref{eq:mnu1}).)  
We may use the above expressions to estimate the CP asymmetries within the context of this model.
Assuming phases of order unity, $m_1^R\sim $10-25~TeV and $M_\nu^{I,II}\sim 0.05$~eV (from Figs.~\ref{fig:3masses}
and \ref{fig:3masses_50}), one obtains the 
estimates $\left|\epsilon_{\nu_{R_1}}^{I}\right|, \left|\epsilon_{\nu_{R_1}}^{II}\right|\sim 10^{-12}$.
Unfortunately, such values are far too small to account for the observed baryon asymmetry of the universe -- one
typically requires $\epsilon_{\nu_{R_1}}$ to be of order $10^{-6}$ or $10^{-7}$~\cite{Aalseth:2002dt,Hambye:2003ka}.
We have computed the asymmetries numerically for the $v_R=20$ and $50$~TeV data sets using the 
original expressions in Eqs.~(\ref{eq:epsIexact})
and (\ref{eq:epsIIexact}), and setting $m_{\Delta_L}=v_R$ for simplicity.  We find numerically
that $\left|\epsilon_{\nu_{R_1}}^{II}\right|\sim 10^{-13}$, with values sometimes of order $10^{-12}$. The asymmetry
$\left|\epsilon_{\nu_{R_1}}^{I}\right|$ is typically of order $10^{-12}$ to $10^{-11}$, but is sometimes
enhanced by one or more
orders of magnitude.\footnote{The approximation $x_j\gg 1$ is not a very good one for this model since
$x_2\simeq 1+{\cal O}(\epsilon)$, as may be inferred from Eqs.~(\ref{eq:A17}) and (\ref{eq:A20}) in the 
Appendix.  The factor $1/(1-x_j)$ in Eq.~(\ref{eq:epsIexact}) diverges for $x_2\to 1$.}
The root cause of the tiny asymmetries is the fact that the 
right-handed scale is so low -- Eqs.~(\ref{eq:epsIapprox}) and (\ref{eq:epsIIapprox})
are both proportional to the lightest right-handed neutrino mass, which is in turn proportional
to the right-handed scale.  If one were to consider a much higher right-handed scale (while
keeping $M_\nu$ fixed at its physical value of approximately $0.05$~eV), one could obtain
asymmetries that are of the correct order of magnitude for leptogenesis.

While leptogenesis would require a much higher right-handed scale than we are considering 
in this work, a low or moderate right-handed scale has the phenomenological
advantage that departures from the SM could be observable 
at upcoming experiments.  
One striking experimental
signature of the LRM would be the production of like-sign leptons
due to the decay of doubly-charged Higgs bosons, $\delta_{L,R}^{\pm\pm}$.
Several authors have investigated
the possibility of producing doubly-charged Higgs bosons at
upcoming collider experiments such as the LHC~\cite{Huitu:1996su,Maalampi:2002vx,Azuelos:2005uc} or
a linear collider~\cite{Godfrey:2001pd,Mukhopadhyaya:2005vf}.  We shall not consider direct Higgs
production further
here, except to note that a lower right-handed scale is obviously 
desirable if one hopes to produce on-shell, doubly-charged Higgs bosons.
The doubly charged Higgs bosons in the LRM also generically contain
lepton flavour violating (LFV) couplings, which are related to the complex symmetric matrix $F$
in Eq.~(\ref{eq:yuk}).  
These couplings lead to 
decays such as $\mu\to 3e$~\cite{Swartz:1989qz}, $\tau\to 3\mu$ and $\tau\to\mu e e$.  
Since the decays
occur at tree-level in the LRM, they may be used to place indirect limits on
various combinations of the LFV couplings and the doubly-charged scalar masses.
We will consider some of the limits for these decays,
as well as branching ratio predictions
for various combinations of LFV couplings and Higgs boson masses.
Other bounds
on the elements of $F$ can also be obtained by considering muonium-antiuonium 
conversion~\cite{Swartz:1989qz,Godfrey:2001pd} and 
$\mu\to e\gamma$~\cite{Mohapatra:1992uu,Barenboim:1996vu,Boyarkin:2004zd}.

Since the $2$-$2$ and $2$-$3$ elements of $F$ are generically rather large in 
our model (see Eq.~(\ref{eq:FGH_eps})), let us first consider the decay $\tau^-\to\mu^+ \mu^-\mu^-$.
Neglecting the muon masses compared to $m_\tau$, we obtain the following expression
for the partial width, 
\begin{eqnarray}
        \Gamma(\tau^-\to\mu^+ \mu^-\mu^-) \simeq \frac{1}{2} \frac{m_\tau}{(32)(192\pi^3)}\left[
	  \left(\frac{m_\tau}{m_{\Delta_L}}\right)^4\left|F_{L23}F_{L22}^*\right|^2+
	  \left(\frac{m_\tau}{m_{\Delta_R}}\right)^4\left|F_{R23}F_{R22}^*\right|^2 \right] ,
	\label{eq:tau3mu}
\end{eqnarray}
where the factor of $1/2$ accounts for the presence of two identical particles in the final state and  
\begin{eqnarray}
        F_{L,R} \equiv V_{L,R}^{\ell T}FV_{L,R}^{\ell } \; . 
	 \label{eq:Frot}
\end{eqnarray}
The matrices $V_{L,R}^\ell$ in the above
expression are those that were used to diagonalize the charged lepton
mass matrix in Eq.~(\ref{eq:diagchargedlep}).  
$V_{L}^\ell$ and $V_{R}^\ell$ both typically involve small mixing angles (see
the discussion in Appendix~\ref{sec:appendix1}).
A cross-term involving $F_L$ and $F_R$ has been dropped
from Eq.~(\ref{eq:tau3mu}) because its leading behaviour is proportional to $m_\mu^3$.
The masses of the doubly-charged Higgs bosons were calculated approximately
in Ref.~\cite{kiers_HiggsLR} for this model.  For our purposes it is
sufficient to keep the leading terms,
\begin{eqnarray}
        m_{\Delta_L}^2 & \simeq & \left(\frac{1}{2}\widetilde{\rho}_3-\widetilde{\rho}_1\right)v_R^2 \; ,\\
	m_{\Delta_R}^2 & \simeq & 2\widetilde{\rho}_2 v_R^2 \; ,
\end{eqnarray}
where the constants $\widetilde{\rho}_1$, $\widetilde{\rho}_2$ and $\widetilde{\rho}_3$ 
are dimensionless coefficients in the Higgs potential
that are generically of order unity in the horizontal symmetry model.

\begin{table*}
\caption{Doubly-charged Higgs masses and corresponding branching ratios for various LFV decays.
As described in the text, we assume that $m_{\Delta_L}=m_{\Delta_R}\equiv m_\Delta$,
that $F_L$ and $F_R$ scale in approximately
the same way as $F$ (see Eq.~(\ref{eq:FGH_eps})) and that the ``order unity coefficients''
in these matrices each take on the value ``3.''  
The entries in the third column give Higgs boson masses that would yield branching
ratios at the current experimental limits.  The
branching ratios for the rare $\tau$ decays are taken
from Ref.~\cite{Yusa:2004gm} and that for $\mu^-\to e^+e^-e^-$ is taken
from Ref.~\cite{PDG2004}.  The 
sixth column gives the branching ratios that would be obtained for the Higgs boson
masses indicated in the fifth column.}
\begin{ruledtabular}
\begin{tabular}{c|c|cc|cc}
	process & matrix elements & $m_{\Delta}$ & exp'l BR ($90\%$ c.l.)  
	  & $m_{\Delta}$ & future BR \\ \colrule
	$\tau^-\to \mu^+\mu^-\mu^-$ & $\left|F_{23}F_{22}^*\right|=(3\epsilon)\times(3)$ & 
         6.2~TeV & $2.0\times 10^{-7}$ & 20~TeV & $2.0\times 10^{-9}$\\
	$\tau^-\to \mu^+e^-\mu^-$ & $\left|F_{23}F_{12}^*\right|=(3\epsilon)\times(3\epsilon^2)$ & 
         2.2~TeV & $2.0\times 10^{-7}$ & 7.0~TeV & $2.0\times 10^{-9}$ \\
	$\tau^-\to e^+\mu^-\mu^-$ & $\left|F_{13}F_{22}^*\right|=(3\epsilon)\times(3)$ & 
         6.2~TeV & $2.0\times 10^{-7}$ & 20~TeV & $2.0\times 10^{-9}$ \\
	$\tau^-\to e^+e^-\mu^-$ & $\left|F_{13}F_{12}^*\right|=(3\epsilon)\times(3\epsilon^2)$ & 
         2.2~TeV & $1.9\times 10^{-7}$ & 7.0~TeV & $2.0\times 10^{-9}$ \\
	$\mu^-\to e^+e^-e^-$ & $\left|F_{12}F_{11}^*\right|=(3\epsilon^2)\times(3\epsilon^4)$ & 
	 10~TeV & $1.0\times 10^{-12}$ & 18 TeV & $1.0\times 10^{-13}$ \\
\end{tabular}
\end{ruledtabular}
\label{tab:LFV}
\end{table*}

The first row of Table~\ref{tab:LFV} gives a few numerical estimates 
for the decay $\tau^-\to\mu^+ \mu^-\mu^-$ assuming, 
for the sake of simplicity, that $m_{\Delta_L}=m_{\Delta_R}\equiv m_{\Delta}$
We also assume that
$F_L$ and $F_R$ both scale in the same way as $F$ (see Eq.~(\ref{eq:FGH_eps}))
and that the ``order unity'' coefficients
in the elements of $F_L$ and $F_R$ all have a magnitude of ``3.''  Under these
assumptions, the left- and right-handed contributions to Eq.~(\ref{eq:tau3mu}) are equal.
Results are also included for several other LFV decay modes.  The expressions for
those decays are similar to Eq.~(\ref{eq:tau3mu}), with appropriate changes in the
matrix elements of $F_L$ and $F_R$\footnote{Interestingly, the Feynman rule for the 
vertex $\ell_{iL,R}\ell_{jL,R}\delta_{L,R}^{++}$ is
proportional to $F_{L,Rij}=F_{L,Rji}$ (no factor of ``$\frac{1}{2}$''), both for the
case $i=j$ and the case $i\ne j$.  In both cases there are two distinct contributions
to the amplitude; these contributions are equal and cancel the factor of $\frac{1}{2}$.}
and the deletion of the factor of ``$1/2$'' in cases in which the final state does not
contain identical particles.
The entries in the fourth column of the table give upper limits on the
experimental branching ratios at the $90\%$~c.l.  The third column lists the Higgs boson
masses that would yield branching ratios right at the experimental limits.  
The decay $\mu\to 3e$ currently 
gives the furthest reach in terms of the doubly-charged Higgs boson mass, even though
$F_{11}$ is generically the smallest element of $F$ in this model.
The fifth and sixth columns
list some representative values for slightly larger Higgs boson masses and what the corresponding
branching ratios would be.  
For the rare $\mu$ decay we have assumed an order of magnitude improvement in the sensitivity.
For the rare $\tau$ decays we have assumed a branching ratio
of $2\times 10^{-9}$, consistent with the ``several times $10^{-9}$'' sensitivity possible
at a Super B factory~\cite{Akeroyd:2004mj}.
A Super B Factory could probe doubly-charged Higgs boson masses at the 20~TeV level.

\section{Discussion and Conclusions}
\label{sec:conclusions}

In this work we have studied the lepton sector of a Left-Right Model with a low right-handed symmetry breaking scale. 
Such a model can be made phenomenologically viable if the underlying $SU(2)_L\times SU(2)_R \times U(1)$ 
gauge symmetry of the theory is supplemented by a $U(1)$ horizontal symmetry to suppress relevant Yukawa 
couplings. We have analyzed the parameter space of this model numerically with the help of a Monte-Carlo 
approach. The model is able to reproduce the main features 
of the lepton mass spectrum and is able to accomodate experimental constraints on the mixing angles
and mass-squared differences of light neutrinos.

We have also discussed other phenomenological applications of this model, such as lepton-flavor-violating
transitions, which occur due to the presence of doubly-charged Higgs bosons in the theory. These transitions produce the 
most striking experimental signatures of the model. We have considered LFV decays of the type 
$\tau^\pm \to \mu^\mp \mu^\pm \mu^\pm$, which constrain both the masses and couplings of these Higgs bosons.
While the currently-available experimental bounds
on such decays probe the mass ranges of a few TeV for the doubly charged Higgs bosons, a two-order of magnitude 
improvement in the experimental bound for the branching ratios of $\tau^\pm \to \mu^\mp \mu^\pm \mu^\pm$
and $\tau^\pm \to e^\mp \mu^\pm \mu^\pm$
would probe the relevant mass scale of tens of TeV. 
We have noted that a LRM with such a low right-handed symmetry breaking scale 
does not accomodate the required CP-violating 
asymmetries needed for generating the baryon asymmetry of the universe via leptogenesis.
Leptogenesis would generally require a
much higher right-handed scale.  We have also calculated the effective mass for
neutrinoless double beta decay for this model.  The resulting values are probably
beyond the sensitivity of experiments of the near future.

\begin{acknowledgments}
K.K. thanks M.-C. Chen and H. Logan for helpful communication.
The work of K.K.and M.A.~was supported in part by the 
U.S.\ National Science Foundation under Grant PHY--0301964.
M.A. and D.S. were also supported in part by Research Corporation
and by the Science Research Training Program at Taylor University.
A.P.~was supported in part by the U.S.\ National Science Foundation under Grant 
PHY--0244853, and by the U.S.\ Department of Energy under Contract DE-FG02-96ER41005.
The work of A.S. was supported in part by DOE contract No. DE-FG02-04ER41291(BNL).
\end{acknowledgments}


\appendix*

\section{Derivation of some approximate results}
In the first part of the Appendix we derive some approximate results 
for right-handed neutrinos.  Following that we comment
on the possibility of obtaining large mixing angles in this model.

\subsection{Approximate relations for right-handed neutrinos}
\label{sec:appendix1}

As is clear from Fig.~\ref{fig:angles_Delm}, the mixing angles for the right-handed
MNS matrix are such that $\sin^2\theta_{12}^R\sim 0.5$, $\sin^2\theta_{23}^R\sim 1$ and
$\sin^2\theta_{13}^R\sim 0$.  In this subsection of the Appendix we outline an approximate calculation
of the three mixing angles, showing that, indeed,
\begin{eqnarray}
	\sin^2\theta_{12}^R &\simeq & 0.5\left(1-{\cal O}(\epsilon)\right) \; , \nonumber \\
	\sin^2\theta_{23}^R &\simeq & 1-{\cal O}(\epsilon^2) \; ,
		\label{eq:app_sinsq} \\
	\sin^2\theta_{13}^R &\simeq & {\cal O}(\epsilon^4) \; . \nonumber
\end{eqnarray}
More precise expressions for the right-handed mixing
angles may be found below.  We also determine
approximate analytical expressions for the heavy neutrino masses.

To determine the right-handed MNS matrix, one must first diagonalize the charged lepton mass
matrix, $M_\ell$, and the mass matrix for heavy (right-handed) neutrinos, $M_{RR}$.
Diagonalization of these two matrices yields the unitary matrices $V_L^\ell$, $V_R^\ell$
and $V_R^\nu$ (see Eqs.~(\ref{eq:diagchargedlep}) and (\ref{eq:mrdiag})).
The latter two of these are used to form the right-handed MNS matrix, as 
seen in Eq.~(\ref{eq:MNStilde2}),
\begin{eqnarray}
	\widetilde{\cal U}_R^{MNS} &=& V_R^{\ell\dagger}V_R^\nu \;  . \nonumber
\end{eqnarray}
The ``tilde'' indicates that the MNS matrix still needs to be rephased to bring
it into the usual form -- see Eqs.~(\ref{eq:ULMNS_rephased}) and (\ref{eq:URMNS_rephased}).
Our approximate calculation makes use of the fact that we have two small quantities with which
to perform an expansion.  The first, $\epsilon$, is the parameter that breaks the horizontal
symmetry in our model.  Although this parameter is not particularly small (we take
$\epsilon=0.3$ throughout this paper), it still does allow for some progress.
The second small parameter in our calculation is the ratio of the bidoublet Higgs VEVs,
assumed to be $k_2/k_1=3/181\sim\epsilon^4$.

Consider first the charged lepton mass matrix, given in Eq.~(\ref{eq:mlep}).  
Taking into account the small value for the ratio $k_2/k_1$ and the scaling
of the Yukawa matrices in (\ref{eq:FGH_eps}), it is clear that $M_\ell$
is dominated by the term proportional to $k_1$; i.e.,
\begin{eqnarray}
	M_\ell \simeq \frac{1}{\sqrt{2}}H k_1 \; .
		\label{eq:mlep_approx}
\end{eqnarray}
(The term proportional to $k_2$ is suppressed by an overall factor of approximately
$\epsilon^8$.) 
Since $H$ is Hermitian and $k_1$ is real, $M_\ell$ is also Hermitian in
this approximation.  Thus, to a good approximation, $M_\ell$ may be diagonalized
by a single unitary matrix (c.f. the general biunitary transformation
shown in Eq.~(\ref{eq:diagchargedlep})).  To the extent that $M_\ell$ is Hermitian,
the resulting diagonalized mass matrix has real eigenvalues, although some of them
would possibly be negative.  A diagonal sign matrix ($\pm 1$ along the diagonal) can then
be used to correct the signs on the masses.  Thus, to a good approximation, $V_R^\ell$ and
$V_L^\ell$ in Eq.~(\ref{eq:diagchargedlep}) are related,
\begin{eqnarray}
	V_L^{\ell}\simeq V_R^\ell A_\ell^\textrm{\scriptsize{sign}}\; ,
		\label{eq:VRVL_approx}
\end{eqnarray}
where $A_\ell^\textrm{\scriptsize{sign}}$ is a sign matrix used to make the charged
lepton masses positive, and we have
\begin{eqnarray}
	M_\ell^\textrm{\scriptsize diag} \simeq A_\ell^\textrm{\scriptsize{sign}} V_R^{\ell\dagger}
		\left(\frac{1}{\sqrt{2}}H k_1\right) V_R^\ell \; .
\end{eqnarray}
It is convenient to parameterize $V_R^\ell$ as a product of 
three $2\times 2$ unitary rotations, as in Eq.~(\ref{eq:U_zero_mns}).  For our purposes
we define
\begin{eqnarray}
	V_R^\ell = {\cal A}_\ell \; {\cal U}^{(0)}
		(\theta_{12}^\ell,\theta_{23}^\ell,\theta_{13}^\ell,\delta^\ell) \; ,
	\label{eq:V_R_approx}
\end{eqnarray}
where ${\cal A}_\ell$ is a diagonal phase matrix.  The diagonalization
proceeds in the following order: (a) diagonal phase rotation, (b) orthogonal $2$-$3$ rotation, 
(c) unitary $1$-$3$ rotation (including the phase $\delta^\ell$), (d) orthogonal $1$-$2$ rotation.
The form of the matrix $H$,
\begin{eqnarray}
	H\sim \left(\begin{array}{ccc}
		\epsilon^{10} & \epsilon^{8} & \epsilon^{7} \\
		\epsilon^{8} & \epsilon^{6} & \epsilon^5 \\
		\epsilon^{7} & \epsilon^5 & \epsilon^4 \\
		 \end{array}\right) \; ,
\end{eqnarray}
allows us to treat all of the $\theta_{ij}^\ell$ in Eq.~(\ref{eq:V_R_approx}) as small quantities.
The approximate diagonalization yields $\theta_{12}^\ell\sim{\cal O}(\epsilon^2)$,
$\theta_{23}^\ell\sim{\cal O}(\epsilon)$ and $\theta_{13}^\ell\sim{\cal O}(\epsilon^3)$ 
(we do not give the analytical expressions here).

A similar procedure may be followed to diagonalize the matrix $M_{RR}=F v_R/\sqrt{2}$.  Since $M_{RR}$
is complex-symmetrix (not Hermitian), the diagonalization is performed using a unitary
matrix and its transpose (rather than a unitary matrix and its Hermitian conjugate),
\begin{eqnarray}
	M_R^\textrm{\scriptsize diag} &=&
		V_R^{\nu T}\left(\frac{1}{\sqrt{2}}F v_R \right)V_R^\nu \; .
\end{eqnarray}
We parameterize $V_R^\nu$ as
\begin{eqnarray}
	V_R^\nu = {\cal A}_\nu \; {\cal U}^{(0)}
		(\theta_{12}^\nu,\theta_{23}^\nu,\theta_{13}^\nu,\delta^\nu) {\cal B}_\nu\; ,
	\label{eq:V_Rnu_approx}
\end{eqnarray}
where ${\cal A}_\nu$ and ${\cal B}_\nu$ are both diagonal phase matrices.
The form of the complex-symmetric Yukawa matrix $F$,
\begin{eqnarray}
	F\sim \left(\begin{array}{ccc}
		\epsilon^4 & \epsilon^2 & \epsilon \\
		\epsilon^2 & 1 & \epsilon \\
		\epsilon & \epsilon & \epsilon^2 \\
		 \end{array}\right) ,
\end{eqnarray}
yields some clues as to how to proceed with the diagonalization.
First of all, we wish to order the eigenvalues in ascending
order (by magnitude), so the $2$-$2$ element needs to be moved to 
the $3$-$3$ location.  This is accomplished using the
$2$-$3$ rotation that occurs immediately following the initial phase rotation.
Since $\theta_{23}^\nu$ is evidently close to $\pi/2$, we take 
$\cos\theta_{23}^\nu$ to be a small quantity for the purpose
of our approximate calculation.  $\theta_{13}^\nu$ is similarly
regarded as a small quantity.  Performing these first three rotations
(the diagonal phase rotation associated
with ${\cal A}_\nu$, the $2$-$3$ rotation and the $1$-$3$ rotation) yields 
the following partly diagonalized mass matrix, 
\begin{eqnarray}
	\left. M_{RR}\right|_{23,13} \simeq \frac{v_r}{\sqrt{2}}
		\left(\begin{array}{ccc}
		{\cal O}(\epsilon^4) & {\cal O}(\epsilon) & 0 \\
		{\cal O}(\epsilon)  & {\cal O}(\epsilon^2) & 0 \\
		0 & 0 & {\cal O}(1) \\
		 \end{array}\right) \; .
\end{eqnarray}
Clearly the remaining $1$-$2$ rotation must be ``large.''  
Neglecting the $1$-$1$ element in the above
expression, it is straightforward to determine an approximate expression for $\sin\theta_{12}^\nu$
in terms of the $1$-$2$ and $2$-$2$ elements and to verify that $\theta_{12}^\nu \sim \pi/4$.

Combining the approximate expressions obtained for $V_R^\ell$ and $V_R^\nu$, we may finally
determine an approximate expression for the right-handed MNS matrix.  Expanding the expression
in terms of $\epsilon$, we obtain the following approximate
relation for $\sin\theta_{12}^R$,
\begin{eqnarray}
	\sin\theta_{12}^R \simeq \sin\theta_{12}^\nu \simeq \frac{\sqrt{2}|a|}
		{\left[\sqrt{4|a|^2+|b|^2}\left(|b|+\sqrt{4|a|^2+|b|^2}\right)\right]^{1/2}} \; ,
\end{eqnarray}
where 
\begin{eqnarray}
	a= F_{13} ={\cal O}(\epsilon)
\end{eqnarray}
and
\begin{eqnarray}
	b = F_{33}-\frac{F_{23}^2}{F_{22}} ={\cal O}(\epsilon^2) \; .
\end{eqnarray}
Then,
\begin{eqnarray}
	\sin\theta_{12}^R=\frac{1}{\sqrt{2}}\left(1-{\cal O}(\epsilon)\right) \; .
\end{eqnarray}
For the other two mixing angles we obtain
\begin{eqnarray}
	\cos\theta_{23}^R & \simeq & \left|\frac{F_{23}}{F_{22}}+\frac{H_{23}}{H_{33}}\right|
		={\cal O}(\epsilon) \; , \\
	\sin\theta_{13}^R & \simeq & \left|\frac{F_{12}^*}{F_{22}}+\frac{F_{13}^*F_{23}}{F_{22}^2}
		-\frac{\left(H_{12}H_{33}-H_{13}H_{23}^*\right)}
		{\left(H_{22}H_{33}-\left|H_{23}\right|^2\right)}\right|={\cal O}(\epsilon^2) \; .
\end{eqnarray}
(Recall that $F_{22}$, $H_{12}$, $H_{13}$, $H_{33}$ and $H_{22}$ are all taken to be real in this work,
as discussed
in Sec.~\ref{sec:simp_Yuk}.)  Note that the $2$-$3$ and $1$-$3$ mixing angles receive
comparable contributions from the diagonalizations of $M_\ell$ and $M_{RR}$, while the
$1$-$2$ mixing angle is determined almost exclusively by the diagonalization of $M_{RR}$.
The above approximate expressions for the three right-handed mixing angles agree relatively well
with the results obtained by performing the diagonalizations numerically, except for cases in which
the approximations break down (such as when a denominator in one of the expressions
involved is accidentally close to zero).

Having performed an approximate diagonalization of $M_{RR}$ to obtain the mixing angles, it is also
straightforward to obtain approximate expressions for the three heavy neutrino masses.  We find
\begin{eqnarray}
	m_1^R & \simeq & \frac{v_R}{2\sqrt{2}}\left[\sqrt{4|a|^2+|b|^2}-|b|\right] ={\cal O}(\epsilon v_R)\; , 
	  \label{eq:A17}\\
	m_2^R & \simeq & \frac{v_R}{2\sqrt{2}}\left[\sqrt{4|a|^2+|b|^2}+|b|\right] ={\cal O}(\epsilon v_R)\; , 
	  \label{eq:A18}\\
	m_3^R & \simeq & \frac{v_R}{\sqrt{2}}F_{22} ={\cal O}(v_R) \; ,
\end{eqnarray}
which leads to the following approximate relation between $m_1^R$ and $m_2^R$ in this model,
\begin{eqnarray}
	m_2^R - m_1^R \simeq \frac{v_r}{\sqrt{2}}
		\left|F_{33}-\frac{F_{23}^2}{F_{22}}\right| ={\cal O}(\epsilon^2v_r) \; .
		\label{eq:A20}
\end{eqnarray}
We may also determine the largest value we might expect for $m_1^R$ in our numerical work (given that 
``order unity'' coefficients are required to have a magnitude between zero and 3).  Setting $b=0$ and
$|a|=3\epsilon$ we obtain
\begin{eqnarray}
	m_{1,\textrm{\scriptsize{max}}}^{R} \simeq \frac{3 \epsilon v_R}{\sqrt{2}} \; .
\end{eqnarray}
For $v_R=20$~TeV, $m_{1,\textrm{\scriptsize{max}}}^{R}\simeq 13$~TeV and for 
$v_R=50$~TeV, $m_{1,\textrm{\scriptsize{max}}}^{R}\simeq 32$~TeV, consistent with 
Figs.~\ref{fig:3masses} and \ref{fig:3masses_50},
respectively.  
The largest values for $m_3^R$ typically occur when $F_{22}\sim 3$, yielding
$m_{3,\textrm{\scriptsize{max}}}^{R}\simeq 42$~TeV for $v_R=20$~TeV and
$m_{3,\textrm{\scriptsize{max}}}^{R}\simeq 110$~TeV for $v_R=50$~TeV.
The approximate expressions for the masses of the three heavy neutrinos agree
relatively well with the values obtained numerically (except for cases
in which the approximations break down, as described above).

The approximate diagonalization procedure also allows us to derive an approximate relation between $\alpha_1^R$
and $\alpha_2^R$, two of the right-handed Majorana phases.  To a good approximation we obtain
$\alpha_2^R\simeq \alpha_1^R+\pi$ (mod $2\pi$), a relation that is evident in Fig.~\ref{fig:phases}.


\subsection{A few comments on obtaining large mixing angles}
\label{sec:appendix2}

The authors of Ref.~\cite{khasanov} noted that it would be difficult
or impossible to obtain large $1$-$2$ mixing in this model.
Looking at Eq.~(\ref{eq:mnu_order_of_mag}), it 
would appear that the $1$-$2$ and
$2$-$3$ angles should indeed be small generically.  Yet our Monte
Carlo algorithm does find order-unity sets of coefficients for $\widetilde{F}$, $\widetilde{G}$ and $\widetilde{H}$ that yield
neutrino masses and mixings consistent with experiment (i.e., with large $1$-$2$ and
$2$-$3$ mixing angles).  
To understand how this can happen, note that the 
diagonalization of the neutrino mass matrix 
may be understood to proceed in several steps, beginning
with a $2$-$3$ rotation, which is followed by a $1$-$3$ rotation and
finally by a $1$-$2$ rotation.\footnote{For simplicity we will
ignore the diagonalization of the charged lepton mass matrix in our considerations here
and consider only the neutrinos' contributions to ${\cal U}_L^{MNS}$
(see Appendix~\ref{sec:appendix1}).}
Experimentally, the $2$-$3$ rotation is large (of order $\pi/4$), the $1$-$3$ rotation is small
and the $1$-$2$ rotation is large (of order $\pi/6$ or $\pi/5)$.
To simplify our discussion, let us consider a case study of a neutrino mass
matrix that was obtained for the case $v_R=20$~TeV.  The magnitudes of the elements
of the mass matrix had the following values after each stage of the diagonalization:
\begin{eqnarray}
	M_\nu \sim \left(\begin{array}{ccc}
		0.0015 & 0.0061 & 0.014 \\
		0.0061 & 0.033 & 0.025 \\
		0.014 & 0.025 & 0.020 \\
		 \end{array}\right)~\mbox{eV} 
	& \stackrel{\textrm{\scriptsize{2-3}}}{\longrightarrow} & 
	       \left(\begin{array}{ccc}
		0.0015 & 0.010 & 0.011 \\
		0.010 & 0.0070 & 0.0021 \\
		0.011 & 0.0021 & 0.052 \\
		 \end{array}\right)~\mbox{eV} \nonumber\\
	& & \nonumber\\
	& \stackrel{\textrm{\scriptsize{1-3}}}{\longrightarrow} & 
	       \left(\begin{array}{ccc}
		0.0038 & 0.011 & 0 \\
		0.011 & 0.0070 & 0 \\
		0 & 0 & 0.054 \\
		 \end{array}\right)~\mbox{eV} \nonumber\\
	& & \nonumber\\
	& \stackrel{\textrm{\scriptsize{1-2}}}{\longrightarrow} & 
	       \left(\begin{array}{ccc}
		0.010 & 0 & 0 \\
		0 & 0.014 & 0 \\
		0 & 0 & 0.054 \\
		 \end{array}\right)~\mbox{eV} \; . \nonumber
\end{eqnarray}
The mixing angles for the diagonalization of this mass matrix
were ``typical;'' i.e., the $2$-$3$ rotation angle was approximately $\pi/(3.6)$,
the $1$-$3$ rotation was small and the $1$-$2$ rotation was approximately
$\pi/(5.6)$.
The first thing that is clear is that the original mass matrix does not
have the hierarchy described in Eq.~(\ref{eq:mnu_order_of_mag}).  In
particular, the $2$-$3$
block is approximately degenerate and is suppressed relative to the ``generic'' expression.
That the $2$-$3$ block is quasi-degenerate is perhaps not a surprise, given that the $2$-$3$ rotation
angle needs to be close to $\pi/4$.  
The elements of the $1$-$2$ block of the original mass matrix have approximately 
the expected magnitudes, with the ratio $\left|M_{\nu 12}/M_{\nu 22}\right|$ being
relatively small.
Following the (large) $2$-$3$ rotation and the (small)
$1$-$3$ rotation, the $1$-$2$ block is in a form suitable for a relatively
large $1$-$2$ rotation.  The neutrino mass matrix in this case study
is fairly typical in that the neutrino mass matrices that 
yield experimentally viable mixing angles tend to have suppressed (and quasi-degenerate) $2$-$3$ blocks
compared to the ``generic'' expectation in Eq.~(\ref{eq:mnu_order_of_mag}).
Some of the suppression is due to a suppression of $F^{-1}$ compared to the
expression in Eq.~(\ref{eq:finv}).


\end{document}